\documentclass[sigconf, nonacm]{acmart}
\pdfoutput=1 % arXiv: build with pdflatex

\usepackage{hyperref}
\usepackage[capitalise]{cleveref}   %
\usepackage{pgfplots}\pgfplotsset{compat=1.18}
\usepgfplotslibrary{groupplots}
\usepackage{siunitx}
\usepackage{tikz}
\usetikzlibrary{arrows.meta, backgrounds, fit, positioning, calc, shapes}
\tikzset{ %
	extends/.style={->, >={Triangle[open, width=0.25cm, length=0.25cm]}}
}
\usepackage{tabularx}
\usepackage{listings}
\usepackage{xcolor}
\usepackage{colortbl}
\usepackage{url}
\usepackage{subcaption}
\usepackage{amssymb}
\usepackage{pifont}
\usepackage[nointegrals]{wasysym}
\usepackage{booktabs}
\usepackage{CJKutf8}   %
\newcommand{\jp}[1]{\begin{CJK}{UTF8}{ipxm}#1\end{CJK}}
\providecommand{\pct}[1]{{\small\textcolor{black!100}{(#1\%)}}}   %
\usepackage{array}

\lstdefinelanguage{Cypher}{
  keywords={MATCH, RETURN, WHERE, CREATE, DELETE, SET, MERGE, OPTIONAL, WITH, AS, AND, OR, NOT, LIMIT, ORDER, BY},
  sensitive=true,
  morecomment=[l]{//},
  morestring=[b]"
}

\usepackage{xcolor}

\usepackage{booktabs,amsmath}

\newbool{showcomments}

   \boolfalse{showcomments}

    \newcommand{\pinaforecomment}[4]{%
    \ifbool{showcomments}{%
        \colorbox{#1}{\textcolor{#4}{\parbox{.8\linewidth}{#2: #3}}}%
    }{}%
}

\AtBeginDocument{%
  }

\newcommand{\mytoolname}{\textit{pykci}}

\newcommand{\benchname}{\textit{Authenti\-City}}

\newcommand{\dsq}[1]{\par\smallskip\noindent\emph{#1}\space}
\newsavebox{\benchtodobox}

\begin{document}

\sloppy

\title{AuthentiCity: A Multi-Source Provenance-Aware Knowledge Graph and Benchmark for 3D City Models}

\author{Huynh Duc An Son Nguyen}
\email{son.nguyen@hcu-hamburg.de}
\orcid{https://orcid.org/0000-0001-8711-1587}
\affiliation{%
  \institution{HafenCity University Hamburg, \\ Computational Methods Lab}
  \streetaddress{Henning-Voscherau-Platz 1}
  \city{Hamburg}
  \state{}
  \country{Germany}
  \postcode{20457}
}

\author{Lukas Arzoumanidis}
\email{lukas.arzou@hcu-hamburg.de}
\orcid{https://orcid.org/0000-0001-6668-1695}
\affiliation{%
  \institution{HafenCity University Hamburg, \\ Computational Methods Lab}
  \streetaddress{Henning-Voscherau-Platz 1}
  \city{Hamburg}
  \state{}
  \country{Germany}
  \postcode{20457}
}

\author{Youness Dehbi}
\email{youness.dehbi@hcu-hamburg.de}
\orcid{https://orcid.org/0000-0003-0133-4099}
\affiliation{%
  \institution{HafenCity University Hamburg, \\ Computational Methods Lab}
  \streetaddress{Henning-Voscherau-Platz 1}
  \city{Hamburg}
  \state{}
  \country{Germany}
  \postcode{20457}
}

\renewcommand{\shortauthors}{Nguyen et al.}

\begin{abstract}
Urban digital twins increasingly combine authoritative, crowd-sourced, machine-learned, and reconstructed data with differing reliability, coverage, and semantics. Yet few urban datasets provide a unified representation that supports multi-source integration, provenance tracking, spatial reasoning, and machine learning. As a result, existing benchmarks rarely evaluate reasoning about source origin, confidence, coverage, and agreement. We present \benchname{}, a multi-source, provenance-aware 3D city knowledge graph spanning five cities across three continents (Hamburg, Helsinki, Zurich, New York, and Tokyo) and comprising 180 GiB, 180M nodes, 220M edges, 1.2B properties, and 3.6M buildings. The labeled property graphs integrate authoritative CityGML features with OpenStreetMap data for all cities, adding roof-material predictions and reconstructed LoD3 geometry for Hamburg, under a provenance model in which derived information never replaces authoritative data. Confidence-weighted edges resolve many-to-many cross-source correspondences, constructing canonical urban entities while preserving traceable links to all contributing evidence. \benchname{} is primarily a data contribution. We introduce two benchmark families to demonstrate the tasks enabled by the representation. The first evaluates natural-language-to-query translation with tasks beyond conventional text-to-SQL and text-to-Cypher benchmarks, including 3D spatial reasoning, provenance-aware filtering, cross-source agreement and disagreement, coverage-aware aggregation, and infeasible-query detection. The second evaluates graph representation learning through multi-source attribute prediction, node classification, and cross-source matching prediction, facilitating comparison of provenance-agnostic and provenance-aware embeddings. Even a strong commercial LLM reaches only 54--69\,\% execution accuracy and a 7B open-weight model 6--19\,\%, and the open-weight model never abstains on an unanswerable question. We release the complete artifact under open licenses with an archival DOI: the five enriched property graphs, loaders, the question suite with gold queries and materialized answers, task splits, an evaluation harness, and a datasheet.
\end{abstract}

\begin{CCSXML}
<ccs2012>
   <concept>
       <concept_id>10002951.10002952.10002953.10010146</concept_id>
       <concept_desc>Information systems~Graph-based database models</concept_desc>
       <concept_significance>500</concept_significance>
       </concept>
 </ccs2012>
\end{CCSXML}

\ccsdesc[500]{Information systems~Graph-based database models}

\keywords{Dataset, Benchmark, CityGML, OSM, Knowledge Graph}

\begin{teaserfigure}
  \includegraphics[width=\textwidth]{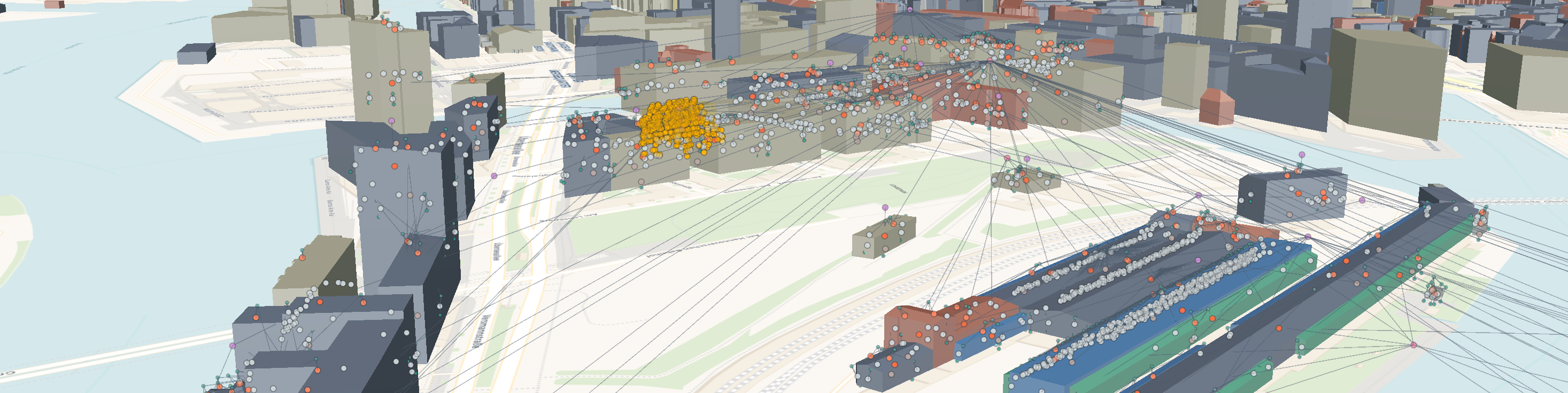} %
  \caption{
  Hamburg, one of five \benchname{} datasets: a provenance-aware 3D knowledge graph integrating CityGML, OpenStreetMap, roof-material predictions (color-coded buildings), and reconstructed LoD3 geometry (orange subgraph). 
  }
  \Description{3D building models colored by predicted roof material, with one reconstructed LoD3 building highlighted in gold, overlaid with a spatially aligned knowledge graph of colored nodes and edges above an OpenStreetMap basemap of the HafenCity district in Hamburg.}
  \label{fig:teaser}
\end{teaserfigure}

\maketitle

\section{Introduction}\label{sec:introduction}

More than half of the world's population lives in cities~\cite{un_urban_2018}, and national mapping agencies increasingly provide authoritative semantic 3D city models at city, regional, and national scales~\cite{nyc_3dbuildingmodel_2016,plateau_3dcitymodel_2021,adv_buildings_2025}. Typically encoded in CityGML~\cite{groeger_citygml2_2012, kolbe_citygml3_2021}, these models provide virtual representations of urban environments and form a key component of urban digital twins (UDTs) for planning, simulation, and decision-making~\cite{ketzler_udtreview_2020, lei_udt_survey_2023}. Yet no single source provides a complete description of a city. Cadastral models offer surveyed geometry and official semantics but often lack use-level detail. Crowd-sourced maps such as OpenStreetMap (OSM) add names, addresses, points of interest, and street networks, but vary in coverage and quality~\cite{biljecki_qualitycrowsourced_2023}. Machine learning can infer attributes such as roof materials from orthophotos~\cite{lukas_roofmaterials, kanna2026semantic}, while photogrammetry provides detailed facade geometry. Integrating these sources therefore requires reasoning over facts that differ in origin, confidence, and coverage.

Existing benchmarks do not directly evaluate this capability. Text-to-query benchmarks such as Spider~\cite{yu_spider_2018}, BIRD~\cite{li_bird_2023}, and CypherBench~\cite{feng_cypherbench_2025} focus on relational or encyclopedic data and do not combine spatial predicates, 3D city geometry, and source provenance. Urban benchmarks such as CityBench~\cite{feng_citybench_2025} and UUKG~\cite{ning_uukg_2023} evaluate urban reasoning or spatiotemporal prediction, but neither exposes a queryable semantic 3D city model nor represents differing trust levels across integrated sources. Consequently, current benchmarks cannot test whether systems select authoritative rather than predicted values, qualify uncertain evidence, or recognize unsupported aggregates arising from incomplete coverage. We refer to these capabilities as \emph{provenance-aware reasoning}.

We introduce \benchname{}, a multi-city knowledge graph (KG) and benchmark for evaluating provenance-aware querying and representation learning over heterogeneous 3D city data. The KG is designed around traceability: authoritative, crowd-sourced, predicted, and reconstructed information are represented as distinct entities and relations. External objects remain separate nodes, fusion edges record matching confidence, and source, confidence, and coverage become explicit inputs to benchmark evaluation.
\benchname{} is constructed with \mytoolname{}~\cite{nguyen_pykci_2026}, an open-source pipeline for mapping CityGML datasets to a compact labeled property graph (LPG) in Neo4j with an R-tree spatial index. On top of the authoritative layer, \benchname{} integrates OSM through confidence-weighted matching edges, attaches machine-learned roof-material predictions, and incorporates reconstructed LoD3 building models.

The dataset follows a two-tier design that balances cross-city comparability with source depth. Tier~1 applies the same CityGML-plus-OSM construction to all cities, enabling a comparable evaluation core. Tier~2 provides a deep-fusion instance for Hamburg, where all four sources are available: CityGML, OSM, ML predictions, and LoD3 reconstructions. The dataset also preserves incomplete coverage rather than masking it. For example, roof-material predictions are available for only about half of Hamburg's buildings, enabling evaluation of whether models distinguish unavailable facts from negative facts and avoid unsupported city-wide conclusions.

We define two complementary benchmark task families. The first evaluates natural-language-to-query translation with gold Cypher over the property graph, covering spatial predicates, cross-source agreement and disagreement, provenance-filtered retrieval, coverage-aware aggregation, and infeasible-question detection (\Cref{subsec:query_tasks}). The second evaluates graph representation learning through attribute imputation, node classification, and matching-link prediction, comparing provenance-agnostic and provenance-aware embeddings (\Cref{subsec:rl_tasks}). Together, these tasks evaluate provenance awareness at both the symbolic and representation-learning levels.

Our contributions are: 
\begin{itemize} 
  \item \textbf{A provenance-preserving, multi-source 3D city dataset.} We release five two-tier city KGs with 180~GiB of data, 180\,million nodes, 220\,million edges, 1.2\,billion properties, and 3.6\,million buildings. The KG integrates authoritative CityGML data, crowd-sourced OSM data, ML-predicted attributes, and reconstructed LoD3 geometry while preserving source provenance through canonical feature nodes and source-specific attachments. The release includes loaders, benchmark splits, a datasheet~\cite{gebru_datasheets_2021}, and an archival DOI. 
  \item \textbf{A benchmark for provenance-aware querying and graph learning.} We provide a text-to-query suite covering spatial, cross-source, coverage-aware, and infeasibility cases, together with a representation-learning suite for attribute imputation, node classification, and matching-link prediction. 
  \item \textbf{Baselines and diagnostic analysis.} 
  We report reference results and a failure decomposition that separates Cypher-generation errors from semantic errors and from failures to abstain.
  For graph learning, we compare provenance-agnostic and provenance-aware variants to measure the value of explicitly representing source information. 
\end{itemize}

\section{Related Work}\label{sec:related_work}

\subsection{Natural-Language-to-Query Benchmarks}

Text-to-SQL is the most mature setting. Spider~\cite{yu_spider_2018} established cross-domain evaluation (\num{10181} questions over 200 databases) but its schemas are small and carry no spatial types. BIRD~\cite{li_bird_2023} scaled to \num{12751} questions over 95 large, noisy databases and introduced the execution-centric evaluation %
we adopt, yet still contains no geospatial reasoning and treats every value as equally trustworthy. Spider~2.0~\cite{li_spider2_2025} adds enterprise-scale realism but remains relational. NL2SQL-BUGs~\cite{liu_nl2sqlbugs_2025} shifts focus to detecting semantically incorrect SQL, which motivates our infeasibility category, and Dr.Spider~\cite{chang_drspider_2023} stresses robustness under perturbations; neither addresses spatial reasoning, provenance, or source disagreement.

For property graphs, the Neo4j Text2Cypher dataset~\cite{ozsoy_text2cypher_2025} aggregates about \num{44000} instances, though many are not grounded in an executable graph. CypherBench~\cite{feng_cypherbench_2025} provides 11 Wikidata-derived graphs with over \num{10000} questions and the execution-accuracy metric we adopt, but its graphs are encyclopedic, with no geometry. Mind the Query~\cite{chauhan_mindthequery_2025} contributes more than \num{27000} validated text-to-Cypher pairs with a rigorous validation pipeline we take as a quality template. All share two limits relevant here: their schemas contain neither spatial geometry nor multiple sources describing the same entity, so the capabilities \benchname{} targets are outside their scope.

\subsection{Urban Benchmarks and Knowledge Graphs}

CityBench~\cite{feng_citybench_2025} evaluates LLMs and VLMs on eight urban tasks across 13 cities but exposes no queryable semantic 3D city model. CityGPT~\cite{feng_citygpt_2025} embeds urban knowledge into the model itself, which does not generalize across cities or updates and gives no auditable grounding. UUKG~\cite{ning_uukg_2023} releases unified urban KGs for spatiotemporal prediction and UrbanKGent~\cite{ning_urbankgent_2024} automates KG construction with LLM agents, but both operate on POI- and region-level entities rather than 3D building models, and neither provides queryable provenance. Surveys of urban KGs and digital twins~\cite{liu_urbankg_2023, wang_urbankgreview_2024, saifwajid_kgudt_2024, akroyd_kgudt_2021} consistently name data fusion as a primary motivation, yet provenance is rarely a first-class queryable component.

Closest to our sources,~\citet{ding_osmgraph_2025} integrate CityGML and OSM into an RDF KG queryable with GeoSPARQL, but the integration is ontology-mediated and not released as a benchmark with tasks, splits, and baselines. KCityChatBot~\cite{liu_kcitychatbot_2025} pairs a CityGML KG with a multi-agent LLM pipeline but provides no reusable benchmark. On the systems side, 3DCityDB~\cite{yao_3dcitydb_2018, yao_3dcitydb5_2025} is the most widely adopted CityGML platform, and Semantic Web representations have been studied extensively~\cite{chadzynski_sem3dcitydb_2021}, but neither natively supports confidence-weighted cross-source correspondences or fact-level provenance, the capabilities central to our tasks (\Cref{subsec:query_tasks}). \benchname{} is complementary to these systems.

\subsection{Multi-Source Integration and Matching}

Fusing crowd-sourced and authoritative geodata requires entity resolution across polygon datasets. Optimal many-to-many polygon matching under the Jaccard measure is NP-hard~\cite{naumann_nnmatching_2024}, with scalable formulations building on tree-constrained bipartite matching~\cite{canzar_treematching_2011, naumann_nnmatching_2025}. Rather than commit to a single set of hard matches, \benchname{} retains the full weighted overlap graph as first-class confidence-annotated edges, letting queries and learning methods resolve correspondences as needed (\Cref{sec:dataset}); the released correspondences are also a resource for polygon-matching research, where ground-truth labels remain scarce.

\subsection{Positioning}

\Cref{tab:gap_matrix} summarizes the gap in existing datasets and benchmarks. Each row represents a strong benchmark within a well-established research area, yet, to the best of our knowledge, no existing benchmark combines these dimensions. \benchname{} addresses this gap. To facilitate comparison with the closest prior work, we adopt evaluation metrics compatible with CypherBench and BIRD, particularly the execution accuracy.

\begin{table}[t]
  \caption{\benchname{} and related datasets and benchmarks. The comparison is
  on capability coverage; see the note below on suite size.}
  \label{tab:gap_matrix}
  \begin{tabular}{lccccc}
    \toprule
    Resource & Spatial & Multi & Prov & Infeas & RL \\
    \midrule
    \multicolumn{6}{@{}l}{\textit{Text-to-query benchmarks}}\\
    \quad Spider~\cite{yu_spider_2018} & \ding{55} & \ding{55} & \ding{55} & \ding{55} & \ding{55} \\
    \quad BIRD~\cite{li_bird_2023} & \ding{55} & \ding{55} & \ding{55} & \ding{55} & \ding{55} \\
    \quad NL2SQL-BUGs~\cite{liu_nl2sqlbugs_2025} & \ding{55} & \ding{55} & \ding{55} & (\ding{51}) & \ding{55} \\
    \quad Text2Cypher~\cite{ozsoy_text2cypher_2025} & \ding{55} & \ding{55} & \ding{55} & \ding{55} & \ding{55} \\
    \quad CypherBench~\cite{feng_cypherbench_2025} & \ding{55} & \ding{55} & \ding{55} & \ding{55} & \ding{55} \\
    \quad Mind the Query~\cite{chauhan_mindthequery_2025} & \ding{55} & \ding{55} & \ding{55} & \ding{55} & \ding{55} \\
    \addlinespace[3pt]
    \multicolumn{6}{@{}l}{\textit{Urban data resources and benchmarks}}\\
    \quad CityBench~\cite{feng_citybench_2025} & (\ding{51}) & \ding{55} & \ding{55} & \ding{55} & \ding{55} \\
    \quad UUKG~\cite{ning_uukg_2023} & (\ding{51}) & \ding{55} & \ding{55} & \ding{55} & \ding{51} \\
    \quad \citet{ding_osmgraph_2025} & \ding{51} & \ding{51} & \ding{55} & \ding{55} & \ding{55} \\
    \midrule
    \benchname{} (+\mytoolname{}~\cite{nguyen_pykci_2026}) & \ding{51} & \ding{51} & \ding{51} & \ding{51} & \ding{51} \\
    \bottomrule
  \end{tabular}

  \smallskip\raggedright\footnotesize
  Spatial: predicates over 2D/3D geometries. Multi: multiple sources adding to the same entities. Prov: explicit provenance representation and provenance-aware evaluation. Infeas: deliberately unanswerable queries. RL: representation learning tasks. Parenthesized (\ding{51}) marks partial or narrower support. 
  The table compares \textbf{capability coverage}, not suite size: \benchname{} releases 1{,}394 executable, gold-verified questions, against 10{,}181 for Spider and more than 10{,}000 for CypherBench over far smaller, non-geometric graphs.
\end{table}

\section{The AuthentiCity Dataset}\label{sec:dataset}

\begin{center}
\fbox{%
  \begin{minipage}{0.93\columnwidth}
    \small\raggedright
    \textbf{The \benchname{} artifact.}\enspace
    Five city-scale knowledge graphs: 180\,GiB, 180M nodes, 220M edges,
    1.2B properties, and 3.6M buildings.
    \par\smallskip
    \textbf{Archive.}\enspace DOI \texttt{10.5281/zenodo.21547211}
    \par\smallskip
    \textbf{Code.}\enspace \url{https://github.com/hcu-cml/authenticity}
    \par\smallskip
    \textbf{Contents.}\enspace Neo4j dump and backend-neutral node/edge
    export, loaders, question suite with gold queries and materialized
    answers, task splits, evaluation harness, and datasheet
    (Appendix~\ref{app:datasheet}).
    \par\smallskip
    \textbf{Licenses.}\enspace Per source (Table~\ref{tab:city_sources});
    OpenStreetMap under ODbL~\cite{opendatacommons_odbl_2026}; our annotations and question
    suite under CC\,BY\,4.0.
  \end{minipage}%
}
\end{center}

This section introduces \benchname{}'s five datasets and its approach to multi-source integration and provenance management. An overview is provided in~\Cref{fig:overview,tab:city_summary}.

\newsavebox{\overviewfigbox}
\begin{figure*}
\centering
\savebox{\overviewfigbox}{%
\colorlet{cAuth}{green!45!black}%
\colorlet{cOsm}{blue!55!black}%
\colorlet{cPred}{orange!75!black}%
\colorlet{cRecon}{violet!65!black}%
\newcommand{\avF}[1]{\textcolor{#1}{\scriptsize\CIRCLE}}%
\newcommand{\avH}[1]{\textcolor{#1}{\scriptsize\LEFTcircle}}%
\newcommand{\avS}[1]{\textcolor{#1}{\scriptsize\Circle}}%
\newcommand{\avN}{\textcolor{gray!60}{--}}%
\newcommand{\famb}[1]{{\setlength{\fboxsep}{1.6pt}\colorbox{black!75}{\textcolor{white}{\scriptsize\bfseries #1}}}}%
\includegraphics{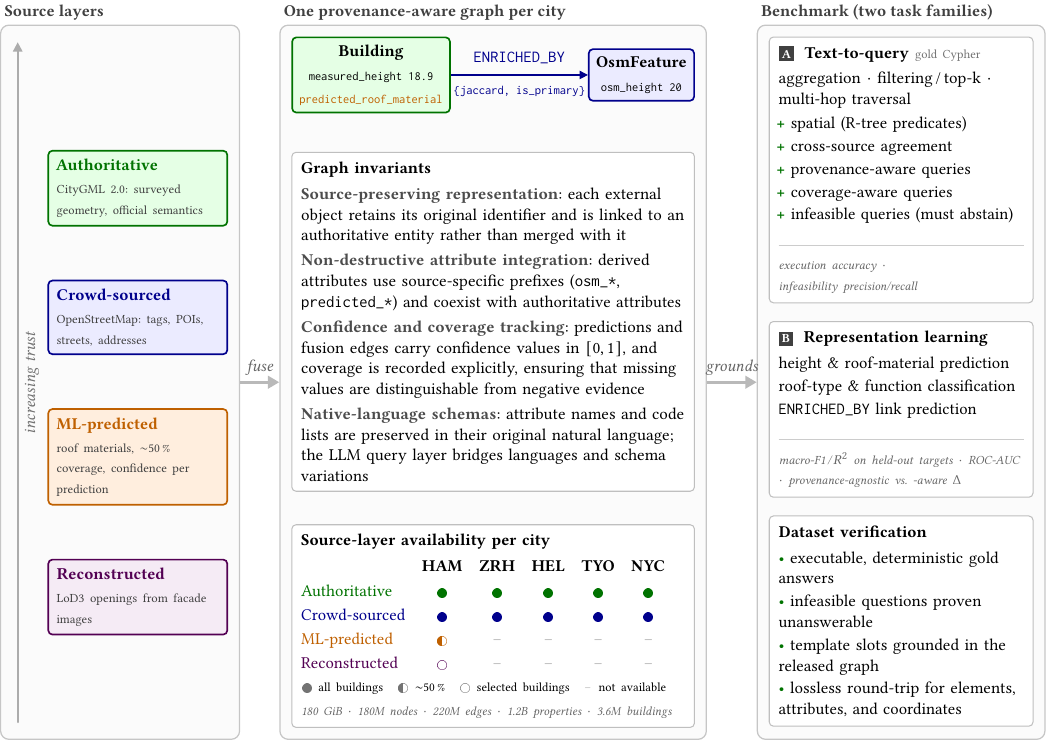}}%
\typeout{OVERVIEWFIG WIDTH: \the\wd\overviewfigbox\space TEXTWIDTH: \the\textwidth}%
\usebox{\overviewfigbox}
\caption{
  Overview of \benchname{}. Four source layers with different trust levels (left) are integrated into a provenance-aware LPG for each city (center). 
  The resulting graphs support two benchmark families with task-specific metrics and dataset-level validation (right). Categories marked (+) are absent from existing text-to-query benchmarks.
}
\Description{Three-panel diagram: four source layers ordered by increasing trust on the left. The fused provenance-aware city graph in the center with a Building to OsmFeature enrichment example, four graph invariants, and a per-city source-layer availability matrix. The two benchmark task families with metrics and dataset verification on the right.}
\label{fig:overview}
\end{figure*}

\begin{table*}[h!]
  \caption{
    Overview of \benchname{}'s five city-scale datasets (full statistics in \Cref{app:dataset_stats}, \Cref{tab:city_sources,tab:city_graph_content}). Hamburg is the Tier~2 deep-fusion instance; the remaining cities form the Tier~1 comparable core. 
    }
  \label{tab:city_summary}
  \setlength{\tabcolsep}{5pt}
  \begin{tabular}{l rrrrr r}
    \toprule
    & Hamburg & Helsinki & Zurich & New York & Tokyo & All 5 \\
    & \small DE & \small FI & \small CH & \small US & \small JP & \small corpus \\
    \midrule
    CityGML / LoD        & 2.0 / LoD2 & 2.0 / LoD2 & 2.0 / LoD2.3 & 1.0/2.0 / LoD1--2 & 2.0 / LoD1--3 & -- \\
    Metric CRS           & EPSG:25832 & EPSG:3879 & EPSG:2056 & EPSG:32618$^\dagger$ & EPSG:6677$^\dagger$ & -- \\
    Buildings            & \num{388267} & \num{2980} & \num{102668} & \num{1083437} & \num{2005762} & \num{3583114} \\
    Nodes                & 17.44\,M & 0.41\,M & 41.51\,M & 45.80\,M & 74.41\,M & 179.57\,M \\
    Edges                & 26.30\,M & 0.62\,M & 45.55\,M & 54.21\,M & 91.95\,M & 218.62\,M \\
    Nodes/bldg.\ (med.)  & 30 & 54 & 253 & 31 & 21 & -- \\
    Thematic keys        & 19 & 49 & 22 & 5 & 20 & -- \\
    OSM-enr.\ (\%)       & 84.9 & 81.1 & 91.2 & 98.7 & 57.8 & 74.1 \\
    Store (GiB)          & 14.8 & 0.3 & 30.2 & 32.4 & 99.2 & 176.9 \\
    Dump (GiB)           & 3.1 & 0.2 & 5.6 & 7.0 & 11.3 & 27.2 \\
    \bottomrule
  \end{tabular}
  
  \smallskip\raggedright\footnotesize
  \emph{Nodes/bldg.} is the median node count per building. \emph{Keys} counts distinct thematic property keys. \emph{OSM-enr.} is the share of buildings matched to at least one OSM feature. \emph{Store}/\emph{Dump} report Neo4j storage size and compressed release size. $^\dagger$ Reprojected at ingest from a non-metric CRS.
\end{table*}

\subsection{Source Layers and the Provenance Spectrum}\label{subsec:provenance_spectrum}

\benchname{} is organized around an explicit \emph{provenance spectrum} (\Cref{fig:overview}): every fact belongs to one of four trust classes recoverable directly from the graph structure rather than external documentation. Three rules enforce this. First, a derived value never overwrites an authoritative one; it is stored under a source-specific property (\texttt{osm\_height} alongside the surveyed \texttt{measured\_height}), so both are comparable in one query. Second, each external object is a separate node that retains its source identity, metadata, and geometry, linked to the authoritative anchor by an explicit edge rather than merged into it. Third, every fusion edge stores its match confidence, so consumers make their own trust decisions rather than inheriting a fixed resolution.

\subsection{Authoritative Layer and Graph Construction}\label{subsec:construction}

The authoritative layer is derived from open-government CityGML~2.0 LoD2 datasets, using Hamburg's state mapping release~\cite{lgv_hhcitygml_2025} with ALKIS cadastral semantics. \mytoolname{} transforms CityGML into a compact Neo4j LPG~\cite{robinson_graphdb_2015}, where semantically meaningful elements become nodes, syntactic wrappers are merged into edges, and coordinates are preserved verbatim. Each top-level feature is additionally registered in an R-tree spatial index~\cite{neo4j_spatial_2025}, enabling semantic traversals and spatial predicates to be combined in a single query. 
The transformation is idempotent (i.e., re-ingest yields an identical graph) and dataset-independent. Mapping details and losslessness evaluation are reported in the companion system paper~\cite{nguyen_pykci_2026}.

CityGML datasets use different coordinate reference systems (CRS), including national metric grids (Hamburg, Zurich, and Helsinki), geographic coordinates (Tokyo), and US survey feet (New York). Rather than enforcing a common CRS, each city is maintained in a local metric CRS to preserve accurate distance and area measurements. 
OSM data are reprojected into the corresponding city-specific CRS before fusion.

\subsection{Identity Resolution and Integrity Check}\label{subsec:fidelity}

Several source datasets violate the assumption that \texttt{gml:id} values are unique, including 100 duplicated identifiers in Zurich and Helsinki and more than \num{975000} (33\,\% of input) ward-boundary identical buildings in Tokyo. 
We resolve these during ingestion without modifying the source files: features sharing an identifier are compared using content and geometry hashes, with identical features merged and distinct features assigned unique identifiers, preventing both erroneous merges and artificial duplication. We further perform pre-ingestion validation to detect identifier, geometry, coordinate, and CRS anomalies, followed by post-ingestion checks of graph integrity, including resolved object counts, identifier uniqueness, provenance completeness, and spatial-index coverage. 

\subsection{Crowd-Sourced Layer: OSM Fusion}\label{subsec:osm_fusion}

OSM provides an independent description of the city. Fusion focuses on footprint correspondence, where one building may map to multiple polygons in either source (1:1, 1:n, n:1, n:m). Candidate pairs $A, B$ come from R-tree intersections and are retained when  $r=\operatorname{area}(A\cap B)/\min(\operatorname{area}(A),\operatorname{area}(B))\ge\tau$.
The overlap coefficient \(r\) serves as the acceptance criterion (\(\tau = 0.3\)). Each accepted edge stores additionally the Jaccard index \(\operatorname{area}(A\cap B)/\operatorname{area}(A\cup B)\). 
Connected components of the bipartite overlap graph define correspondence classes, and component-level rules propagate source-prefixed OSM attributes to matched buildings (\Cref{app:enriched_osm}, \Cref{fig:osm_enrich_example}). Geometry and points of interest remain on OSM nodes.
\Cref{app:enriched_osm} explains how the value of $\tau$ is selected. 
Unlike traditional conflation pipelines, we retain the full weighted overlap graph as \texttt{ENRICHED\_BY} edges rather than enforcing a final matching, which is an NP-hard problem~\cite{naumann_nnmatching_2024}. 
All OSM features are additionally registered in a second, dedicated R-tree layer, so crowd-sourced and authoritative geometry remain independently and jointly queryable.

\subsection{ML-Predicted and Reconstructed Layer}\label{subsec:predicted_layers}

The ML layer adds roof-material classes (concrete, metal, glass, roof tiles, tar paper) predicted from aerial orthophotos~\cite{lukas_roofmaterials}. 
\benchname{} preserves each building's full material distribution, storing every predicted class with its pixel coverage. 
Fusion is a direct ID join, with all values stored under source-prefixed keys. Predictions cover \num{194799} of \num{388267} buildings (50.2\,\%), limited by orthophoto availability. 

The reconstructed layer adds LoD3 building models derived from facade imagery for 17 buildings in Hamburg's HafenCity district. The models contribute 416 windows and 83 doors as explicit \texttt{Opening} nodes, represented by 499 interior rings in wall geometries. %
Corrected facades are linked to their LoD2 buildings via \texttt{HAS\_LOD3\_FACADE} edges as separate nodes in a dedicated spatial layer, never replacing the measured geometry. 

\subsection{Cross-Source Agreement and Conflict}\label{subsec:conflict}

\benchname{}'s datasets are highly diverse 
(see~\Cref{app:dataset_stats},~\Cref{fig:city_stats}). Fusing these urban data sources creates both coverage and redundancy. We analyze geometric correspondences and dual-sourced attributes in Hamburg, Helsinki, Zurich, New York, and Tokyo. 

\paragraph{Geometric disagreement.} Most matched buildings align cleanly: 87.2\,\% of Hamburg's \num{293895} correspondences are 1:1 matches (\Cref{tab:footprint_cases}). 
In 1:n components, one OSM building covers 2.5 CityGML buildings on average; in n:1 components, one CityGML building corresponds to 3.0 OSM buildings. The \num{2363} n:m components motivate retaining the full overlap graph rather than forcing a single matching. Fragmentation also varies by city, from 0.4\,\% in New York to 
9.8\,\% in Tokyo.

\begin{table}[h!]
  \caption{Footprint correspondence between CityGML and OSM (share of
  components per city). Full counts in Table~\ref{tab:footprint_cases}.}
  \label{tab:footprint_cases_lite}
  \begin{tabular}{lrrrrr}
    \toprule
    City & 1:1 & 1:$n$ & $n$:1 & $n$:$m$ & Fragmented \\
    \midrule
    Hamburg  & 87.2 & 6.5 & 5.5 & 0.8 & 6.3 \\
    Helsinki & 84.2 & 6.3 & 6.4 & 3.1 & 9.5 \\
    Zurich   & 90.8 & 5.8 & 2.4 & 0.9 & 3.3 \\
    New York & 99.3 & 0.3 & 0.3 & 0.1 & 0.4 \\
    Tokyo    & 85.9 & 4.3 & 2.3 & 7.5 & 9.8 \\
    \bottomrule
  \end{tabular}
  
  \smallskip\raggedright\footnotesize
  Fragmented comprises the \(n{:}1\) and \(n{:}m\) cases, that is, CityGML buildings split across multiple OSM building footprints. 
\end{table}

\begin{table}[h!]
  \caption{OSM ingestion census.}
  \label{tab:osm_census}
  \begin{tabular}{lrrrrrr}
    \toprule
    City & Ingested & Coverage & Anchored & Standalone \\
    \midrule
    Hamburg  & \num{1188139} & 95.6\,\% & \num{356707} & \num{831432} \\
    Helsinki & \num{55930}   & 89.7\,\% & \num{4281}   & \num{51649} \\
    Zurich   & \num{1860401} & 95.8\,\% & \num{102794} & \num{1757607} \\
    New York & \num{2555468} & 97.6\,\% & \num{1121901} & \num{1433567} \\
    Tokyo    & \num{3830408} & 97.9\,\% & \num{1235353} & \num{2595055} \\
    \bottomrule
  \end{tabular}

  \smallskip\raggedright\footnotesize
  \emph{Ingested} is how many OSM features enter the graph, not including non-network lines like barriers, building-annotation open ways, and man-made linear features.
  \emph{Coverage}: ingested features as a share of all source OSM features in the city's bounding box. 
  \emph{Anchored} vs.\ \emph{Standalone} split the ingested features by whether they attach to a CityGML feature or are kept as net-new nodes.
\end{table}

\paragraph{Attribute agreement and conflict.} Height and storey counts largely agree when they are present (\Cref{tab:dual_attrs}): storeys match exactly for 87.2\,\% of \num{144364} buildings (98.6\,\% within $\pm1$), and the median height difference over \num{3404} buildings is 0.27\,m. Disagreements remain informative, including 219 buildings differing by more than 5\,m. Cross-city behavior differs substantially: Zurich shows much lower height agreement (median $|\Delta h|=2.3$\,m), likely reflecting differing measurement conventions. 

Roof material provides the richest comparison.
Among \num{3641} buildings with both ML and OSM labels, agreement reaches 75.2\,\%.
Agreement is highest for roof tiles (92.6\,\%) and lower for flat-roof materials.
OSM also contributes \num{5059} roof-material values without predictions and 134 values outside the prediction taxonomy.

\subsection{Multilingual Urban Knowledge Graphs}\label{subsec:multilingual}

Helsinki, Hamburg, and especially Tokyo illustrate why natural-language interfaces matter for urban KGs. Tokyo's PLATEAU data uses Japanese attribute names and values, for example \jp{地区計画} (district plan) and \jp{市谷柳町地区}, which \mytoolname{} preserves verbatim as Unicode property keys. Traditionally, querying such data required familiarity with both the local schema and natural language. With LLM-grounded querying, users can ask questions in their own language, 
as shown in this work, making Tokyo's 100\,GiB graph accessible to both local residents and international analysts.

\section{Benchmark A: Natural-Language-to-Query}\label{subsec:query_tasks}

To show that the released graphs can be queried in natural language, and that doing so demands reasoning beyond what conventional text-to-SQL and text-to-Cypher benchmarks test, we define a natural-language-to-query benchmark. Given a natural-language question and the graph schema, a model must either produce an executable Cypher query whose result matches the gold answer or explicitly declare the question infeasible. We evaluate on LPGs by design. While a relational CityGML store can support aggregate, filter, and spatial queries, the benchmark's cross-source agreement, provenance- and confidence-filtered retrieval, and coverage-aware aggregation rely on LPG-native fusion structures absent from relational schemas.

\subsection{Categories}

We design \num{1394} query questions spanning 84 templates and nine categories (\Cref{tab:categories}; full inventory in \Cref{app:questions}), five of which are our contributions. \emph{Spatial} questions, the largest category, cover window, proximity, nearest-neighbor, geometric multi-hop, 3D-structure, and density queries. \emph{Cross-source} questions compare authoritative and crowd-sourced values and ask what the crowd-sourced layer reports where authoritative data is absent, a common case in New York, where \num{97.5}\,\% of buildings have an OSM height but none an authoritative one. \emph{Provenance-filtered} questions constrain answers by source or confidence, including traps (``using only authoritative data\ldots'') where reading a crowd-sourced value yields a plausible but incorrect answer. \emph{Coverage-aware} questions require recognizing that aggregates over partially covered attributes are meaningful only with respect to the covered subset; the flagship trap asks for a city-wide roof-material count, where the gold answer couples count and coverage. \emph{Infeasible} questions have no valid answer and test whether a model declines rather than fabricates a query, each paired with a guard query proving the information absent. 

Two design choices increase difficulty. First, as feasibility depends on city-specific coverage, the same template can be coverage-aware in one city and infeasible in another (e.g., roof material is a coverage question in Hamburg but infeasible elsewhere). Second, ten templates are additionally posed in German and Japanese, including three Tokyo templates whose gold queries filter on Japanese property keys. A thematic subset targets city-specific schema content, including Helsinki's floor area, volume, and construction year; Zurich's data-vintage fields; and Tokyo's Japanese-keyed attributes 
(\Cref{app:thematic}).

\subsection{Construction, Verification, and Metrics}

The benchmark is generated from 84 schema-grounded templates, each paired with a hand-authored gold Cypher query whose slot values are read directly from the released graph. Feasible templates declare the graph tokens they require and instantiate only where those tokens exist; missing attributes are exercised through the infeasible subfamily. Every question passes a five-stage verification gate: gold queries execute, are deterministic, and are non-empty unless permitted; infeasible questions include a guard query; and each rebuild re-materializes all gold answers. This extends the methodology of~\cite{chauhan_mindthequery_2025}. Additionally, \num{282} questions (20.2\,\%) were manually reviewed, revealing a proximity-predicate defect that escaped automated checks and was fixed before release.

The benchmark is split by template into a public development set and a held-out test set with unpublished gold queries. The test set contains unseen templates and holds out 370 of \num{1394} questions (26.5\,\%). We report execution accuracy (result-set equivalence under canonicalization), decomposed into executed-and-correct (EX), executed-but-wrong, errored, and over-refusal. This distinction separates Cypher-generation failures from semantic failures. We also report infeasibility precision, recall, and F1, 
following prior NL-to-Cypher benchmarks~\cite{feng_cypherbench_2025, li_bird_2023}.

\subsection{Setup and Results}\label{subsec:experiments_setup}

We evaluate a local open-weight model (qwen2.5-coder:7b, Ollama \texttt{Q4\_K\_M}) and a commercial frontier model (Claude Sonnet~5, run as a closed-book Claude Code subagent). Both models receive byte-identical, schema-only context with no database or tool access. Outputs are cached and scored offline against the gold queries (examples in~\Cref{app:questions_examples}).
Even the frontier model leaves substantial headroom. Claude achieves 54--69\,\% EX across the five cities (\Cref{tab:results_query}). Nearly all remaining cases are \emph{executed-but-wrong} rather than errored, indicating semantic rather than syntax failures. qwen reaches only 6--19\,\% EX and is dominated by \emph{errored} queries (40--54\,\%), largely Cypher syntax errors in spatial tasks where it hallucinates PostGIS \texttt{ST\_*} functions. This decomposition reveals qualitatively different failure modes that aggregate EX obscures.
\begin{table*}[ht!]
  \caption{
  Task family A results on all five city dev and held-out test splits (\%). The four feasible outcomes sum to 100: EX (correct), Ex-wr (executed but incorrect), Err (execution failure), and O-ref (incorrect refusal). Inf-F1 measures infeasibility detection. 
  }
  \label{tab:results_query}
  \setlength{\tabcolsep}{4pt}
  \begin{tabular}{ll ccccc c ccccc}
    \toprule
    & & \multicolumn{5}{c}{Development set} & & \multicolumn{5}{c}{Held-out test set} \\
    \cmidrule(lr){3-7}\cmidrule(lr){9-13}
    Model & City & EX\,$\uparrow$ & Ex-wr & Err & O-ref\,$\downarrow$ & Inf-F1 & & EX\,$\uparrow$ & Ex-wr & Err & O-ref\,$\downarrow$ & Inf-F1 \\
    \midrule
    Claude   & Hamburg  & 60.7 & 32.6 &  0.0 & 6.7 & 72.0 & & 58.6 & 41.4 &  0.0 & 0.0 & 100.0 \\
    Sonnet 5 & Helsinki & 56.1 & 39.0 &  0.0 & 4.9 & 78.3 & & 42.9 & 57.1 &  0.0 & 0.0 &  66.7 \\
             & Zurich   & 61.5 & 38.5 &  0.0 & 0.0 & 94.7 & & 60.0 & 40.0 &  0.0 & 0.0 & 100.0 \\
             & New York & 69.2 & 28.2 &  0.0 & 2.6 & 90.0 & & 50.0 & 35.7 & 14.3 & 0.0 & 100.0 \\
             & Tokyo    & 53.7 & 37.3 &  7.5 & 1.5 & 90.0 & & 41.7 & 37.5 & 20.8 & 0.0 & 100.0 \\
    \addlinespace
    qwen2.5- & Hamburg  & 19.1 & 36.0 & 44.9 & 0.0 & n/a & & 20.7 & 48.3 & 31.0 & 0.0 & n/a \\
    coder:7b & Helsinki & 14.6 & 45.1 & 40.2 & 0.0 & n/a & & 10.7 & 53.6 & 35.7 & 0.0 & n/a \\
             & Zurich   &  9.2 & 41.5 & 49.2 & 0.0 & n/a & & 15.0 & 40.0 & 45.0 & 0.0 & n/a \\
             & New York & 10.3 & 35.9 & 53.8 & 0.0 & n/a & & 14.3 & 28.6 & 57.1 & 0.0 & n/a \\
             & Tokyo    &  6.0 & 43.3 & 50.7 & 0.0 & n/a & & 12.5 & 33.3 & 54.2 & 0.0 & n/a \\
    \bottomrule
  \end{tabular}

  \smallskip\raggedright\footnotesize
  Inf-F1 is the F1 score for infeasibility detection, where over-refusals count as false positives. qwen2.5-coder never refuses, so Inf-F1 is undefined (n/a) and recall is 0. For qwen2.5-coder, most Err cases are Cypher syntax errors in spatial queries caused by hallucinated PostGIS \texttt{ST\_*} functions. For Claude, most Err cases on the NYC and Tokyo test splits are spatial-query timeouts on the largest graphs (\Cref{subsec:experiments_setup}). Baselines are evaluated on a stratified subsample of the released suite; per-split counts are in the repository. 
\end{table*}
On infeasibility, qwen never refuses (recall~0), generating a query for every unanswerable question, whereas Claude declines appropriately (Inf-F1 72--95 on dev). 
Declines on New York and Tokyo stem primarily from spatial-query timeouts on the largest graphs (1--2M buildings) rather than semantic drift.

\section{Benchmark B: Representation Learning}\label{subsec:rl_tasks}

To demonstrate that the graph supports applications beyond text-to-query, including urban analytics such as energy efficiency~\citep{Yap2025} and urban planning~\citep{LIU2022102936}, as well as provenance-aware tasks such as data fusion, quality assurance~\citep{biljecki2016common, biljecki2021open}, and cross-city transfer that underpin urban digital twins~\citep{abdelrahman2025digitaltwin}, we provide an embedding-based evaluation following the benchmark protocol of~\citet{dwivedi_benchmarkgnn_2023}. It comprises three tasks under spatial-block and cross-city splits, namely \emph{attribute imputation} of ML-predicted roof material (available for only ${\sim}\num{50}\%$ of buildings~\citep{lukas_roofmaterials}) and authoritative building height (T1), \emph{node classification} of administrative building-function and roof-type classes (T2), and \emph{link prediction} of held-out \texttt{ENRICHED\_BY} correspondence edges (T3). Each task is run under a \emph{provenance-agnostic} protocol that hides source distinctions and a \emph{provenance-aware} one that exposes source types and fusion-edge confidences, for example through Jaccard-weighted message passing, and the difference between them measures whether provenance helps. To our knowledge this is the first such evaluation on a real city-scale multi-source graph, although label coverage varies by task, with height spanning all five cities, matching per city, roof type in Hamburg and Helsinki, and building function only Hamburg (Section~\ref{sec:repL:setup}).

\subsection{Provenance-Agnostic vs.\ Provenance-Aware}
\label{sec:repL:prov}
The two protocols differ only in how much of the provenance structure the encoder may use, so that any performance gap is attributable to provenance awareness rather than to model capacity. The \emph{provenance-agnostic} setting flattens the multi-source graph into canonical entities by merging source-specific evidence, removing edge-confidence values, and discarding source labels. The \emph{provenance-aware} encoder instead retains (i) source-typed nodes and relations (CityGML, OSM, ML-derived), (ii) confidence scores as edge weights, and (iii) coverage and agreement node features (source presence, count, and best-match confidence). Specifically, a canonical entity $c$ aggregates evidence from its neighbors $e\in\mathcal N(c)$ as
\begin{equation}
\mathbf h_c' \;=\; \sigma\Big(\mathbf W_{\text{self}}\,\mathbf h_c
\;+\!\!\sum_{e\in\mathcal N(c)}\!\frac{w_{ce}}{\textstyle\sum_{e'} w_{ce'}}\;
\mathbf W_{s(e)}\,\mathbf h_e\Big),
\label{eq:confagg}
\end{equation}
where $w_{ce}\in[0,1]$ is the correspondence confidence (the Jaccard overlap stored on the CityGML--OSM edges, normalized per target node) and $\mathbf W_{s(e)}$ is a source-specific projection, instantiated as one relation-specific weight matrix per typed relation. The agnostic variant sets $w_{ce}=1$ and $\mathbf W_{s(e)}=\mathbf W$, recovering mean-pooled GraphSAGE~\citep{hamilton2017graphsage}.

\subsection{Experimental Setup}
\label{sec:repL:setup}
\paragraph{Models.} At full scale we evaluate an attribute-only MLP, provenance-agnostic GraphSAGE~\cite{hamilton2017graphsage} and GAT~\cite{velickovic2018gat}, source-typed R-GCN~\cite{schlichtkrull2018rgcn}, HAN~\cite{wang2019han}, and HGT~\cite{hu2020hgt}, the confidence-weighted encoder of \cref{eq:confagg}, self-supervised DGI~\cite{velickovic2019dgi}, and non-learned spatial-distance and attribute-similarity baselines for T3. The shallow node2vec~\cite{grover2016node2vec} and metapath2vec~\cite{dong2017metapath2vec} baselines are transductive and cannot embed unseen entities, so we exclude them from the full-scale comparison.

\paragraph{Splits and protocol.} The \emph{spatial} split holds out entire spatial tiles, and because a single hold-out is high-variance at city scale we use \emph{spatial $K$-fold cross-validation} ($K{=}5$) with out-of-fold predictions pooled so every entity is tested once. The \emph{cross-city} split is leave-one-city-out. Shallow methods are trained unsupervised, frozen, and probed, whereas GNNs are trained end-to-end. We report mean$\pm$std over \num{3} seeds for the survey (\Cref{tab:repL-main}) and over \num{10} seeds for the cross-city result. For the focused provenance-agnostic versus provenance-aware paired comparisons and the confidence-weighting ablation, we repeat over \num{10} seeds and pair runs by seed, reporting a paired $t$-test~\citep{gosset1908}, the Wilcoxon signed-rank test~\citep{wilcoxon1945}, Cohen's $d_z$~\citep{lakens2013}, and a 95\% confidence interval on the per-seed difference, and we treat a comparison as a null when the interval contains zero or the effect size is negligible. Provenance-agnostic and aware variants share backbone, depth, and budget (hidden dimension \num{64}, \num{2} message-passing layers, dropout \num{0.3}, Adam at learning rate \num{0.01}, \num{150} epochs), and are trained in PyTorch Geometric on an Nvidia RTX PRO~6000 (96\,GB). To prevent leakage, each predicted attribute is removed from the inputs, and a dedicated ablation additionally removes its correlated counterpart, such as storey count when predicting height. Data are streamed from Neo4j, with the loader handling missing node types, features, and labels so the same code runs unchanged across cities.

\subsection{Experimental Results}
\label{sec:repL:results}
A single-city subset (Hamburg, $n=81$ buildings) first validated the pipeline and previewed both headline findings, that topology adds signal beyond attributes (with the collinear storey feature withheld, the attribute probe falls to $R^2{=}\num{-0.86}$ while GraphSAGE recovers $\num{0.52}$) and that provenance-awareness gives no measurable single-city benefit, including no gain on the seven buildings with conflicting OSM/CityGML roof evidence.

\paragraph{Full run (Hamburg, $n{=}388$k buildings).}
Within a single city, the aware encoder's advantage over the agnostic baseline is statistically consistent but negligible in magnitude. Over \num{10} seeds it improves T1 height ($\num{0.729}\rightarrow\num{0.733}$), T2 roof type ($\num{0.556}\rightarrow\num{0.558}$), and T2 building function ($\num{0.464}\rightarrow\num{0.474}$), each significant under a paired $t$-test ($p<\num{0.02}$) yet at most \num{0.010} in absolute terms (cf.~\Cref{tab:repL-main}, \num{3}-seed run). R-GCN, source-typed but not confidence-weighted, stays within about \num{0.02} of the full encoder on all three tasks, which points to source typing rather than confidence weighting as the origin of even this small effect. We therefore state a falsifiable hypothesis: source typing and confidence weighting are separable, and confidence weighting is redundant with source typing in-distribution but not under distribution shift. We test it with an ablation that removes only the confidence weights, leaving source typing, coverage features, and architecture unchanged. In-distribution, removing confidence weighting changes every task by at most \num{0.001}, with no significant effect on height ($p=\num{0.21}$) or building function ($p=\num{0.59}$) and only a negligible \num{0.0006} on roof type ($p=\num{0.02}$), confirming that confidence weighting is redundant with source typing in-distribution. Whether it also contributes under distribution shift, where the aware encoder's cross-city gains appear (\Cref{tab:repL-crosscity}), requires the same ablation under leave-one-city-out and remains future work. 

In the T3 matching embedding space (\cref{fig:prov-tsne}), the aware encoder separates OSM-matched buildings into a distinct region where the agnostic encoder does not, and the ML- and OSM-coverage gaps co-locate, indicating that the two missingness patterns are correlated rather than independent, a restructuring the supervised metrics alone do not reveal. Under leave-one-city-out on T1 height the aware encoder improves on four of the five held-out cities (\cref{tab:repL-crosscity}), with Hamburg rising from \num{0.270} to \num{0.548} and its standard deviation dropping from \num{0.170} to \num{0.071}, plus gains on Zurich, New York City, and Tokyo, and Helsinki tied within noise. Roof-type transfer, possible only on the Hamburg--Helsinki pair, shows no meaningful gap between the two encoders (\Cref{tab:repL-crosscity-roof}). For T3 matching, every learned encoder falls far short of non-learned baselines, for a structural reason. \texttt{ENRICHED\_BY} correspondences are defined by footprint overlap (median Jaccard \num{0.842}), so on the default nearest-neighbor negatives a spatial-distance rule reaches ROC-AUC \num{0.95}--\num{0.996} while the encoders reach at most \num{0.57}. On the ambiguous n:m subset, where a building overlaps several OSM candidates and distance is uninformative, the distance rule falls to \num{0.75} AUC, a trivial attribute rule (building height versus OSM levels) still reaches \num{0.94}, and every learned encoder collapses to chance (\num{0.47}--\num{0.49} AUC on \num{18234} Hamburg cases). T3 is therefore not a coordinate lookup, but it exposes a concrete gap, namely that current encoders exploit geometry and ignore the cross-source attribute signal that actually disambiguates correspondences. We provide the distance and attribute heuristics as reference baselines and pose attribute-aware n:m matching as an open challenge.

\begin{table}[t]
  \centering
  \caption{Full-scale (single-city) Hamburg results (388k buildings, spatial cross-validation, mean$\pm$std over \num{3} seeds). Height retains the collinear storey count.}
  \label{tab:repL-main}
  \small
  \setlength{\tabcolsep}{4pt}
  \begin{tabular}{@{}lccc@{}}
    \toprule
    & Roof type & Bldg function & Height\\
    Model & macro-F1 $\uparrow$ & macro-F1 $\uparrow$ & $R^2$ $\uparrow$\\
    \midrule
    MLP (attr-only)             & $0.504\pm.000$ & $0.329\pm.000$ & $0.666\pm.000$\\
    DGI (self-supervised)       & $0.516\pm.007$ & $0.409\pm.002$ & $0.683\pm.002$\\
    GraphSAGE (agnostic)        & $0.556\pm.000$ & $0.462\pm.000$ & $0.730\pm.002$\\
    GAT (agnostic)              & $0.531\pm.000$ & $0.365\pm.001$ & $0.676\pm.001$\\
    \multicolumn{4}{@{}l}{\emph{source-typed, not confidence-weighted}}\\
    HGT                         & $0.541\pm.002$ & $0.393\pm.007$ & $0.714\pm.005$\\
    HAN                         & $0.460\pm.003$ & $0.208\pm.004$ & $0.416\pm.026$\\
    R-GCN                       & $0.558\pm.000$ & $0.470\pm.003$ & $0.713\pm.004$\\
    Conf.-weighted GNN (aware)  & $\mathbf{0.559\pm.001}$ & $\mathbf{0.476\pm.001}$ & $\mathbf{0.732\pm.000}$\\
    \bottomrule
  \end{tabular}
\end{table}

\begin{table}[t]
  \centering
  \caption{T1 cross-city height transfer, leave-one-city-out ($R^2$, mean$\pm$std over \textbf{10} seeds, best cross-city result per row in bold), where \emph{own-city} is the single-city reference. The \emph{probe} is a no-graph baseline pooled over the training cities and is not comparable to the single-city attribute probe of \Cref{tab:repL-main}.}
  \label{tab:repL-crosscity}
  \small
  \setlength{\tabcolsep}{4pt}
  \begin{tabular}{@{}lrcccc@{}}
    \toprule
    Held-out & $n$ & probe & agnostic & aware & own-city\\
    \midrule
    Hamburg  & 388{,}267     & $-0.145$ & $0.270\pm.170$ & $\mathbf{0.548\pm.071}$ & 0.733\\
    Helsinki & 2{,}980       & $-0.000$ & $\mathbf{0.516\pm.016}$ & $0.500\pm.029$ & 0.461\\
    NYC      & 1{,}083{,}437 & $\mathbf{0.258}$ & $0.167\pm.034$ & $0.188\pm.024$ & 0.407\\
    Tokyo    & 2{,}005{,}762 & $-0.484$ & $0.011\pm.130$ & $\mathbf{0.082\pm.063}$ & 0.399\\
    Zurich   & 102{,}668     & $-0.011$ & $0.581\pm.035$ & $\mathbf{0.601\pm.037}$ & 0.269\\
    \bottomrule
  \end{tabular}
\end{table}

\paragraph{Cross-source completion and auxiliary targets.} Unlike the three survey tasks, roof-material imputation, which predicts the ${\sim}\num{50}\%$ of Hamburg buildings the external ML model did not label, gives the aware encoder its clearest single-city edge (macro-F1 $\num{0.283}\rightarrow\num{0.316}$, concentrated in the \texttt{concrete} class). A circularity check confirms this is structural rather than leakage, as only \num{3.5}\% of labeled buildings carry a matched OSM roof-material tag.

\section{Conclusion}\label{sec:conclusion}

We presented \benchname{}, a multi-source, provenance-aware 3D city knowledge graph and a two-family benchmark on top of it. The dataset makes origin, confidence, and coverage of urban facts explicit and queryable across authoritative, crowd-sourced, ML-predicted, and reconstructed layers. The benchmark evaluates capabilities that are not captured by existing text-to-query or urban reasoning benchmarks, including spatial reasoning over indexed 3D geometry, cross-source comparison, coverage-aware aggregation, infeasibility detection, and provenance-aware representation learning. We release the complete artifact under open licenses, including the enriched LPGs, question suite, gold queries, data splits, loaders, evaluation harness, and datasheet, together with an archival DOI. We hope \benchname{} will serve both as a rigorous testbed for evaluating query-generation models on realistic, heterogeneous urban data and as a foundation for developing provenance-aware embedding methods, an important gap highlighted by our baseline results. 
The implementation, documentation, and additional resources associated with this project are publicly available at \url{https://github.com/hcu-cml/authenticity}.

\bibliographystyle{ACM-Reference-Format}
\bibliography{sample-base}

\appendix

\begin{figure}[h!]
  \centering
  \begin{subfigure}{0.49\linewidth}
    \includegraphics[width=\linewidth]{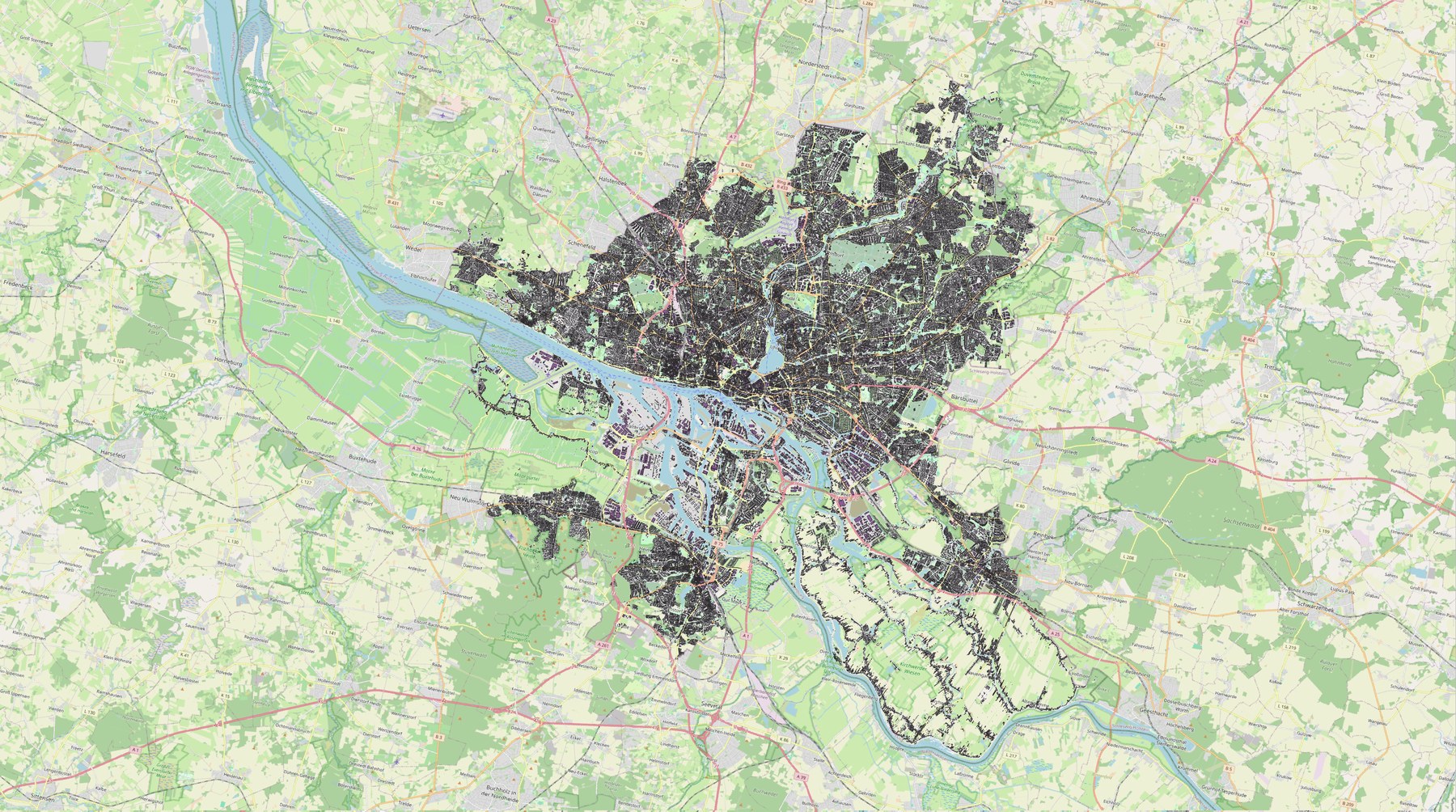}
    \caption{Hamburg, 1:84,460}
  \end{subfigure}\hfill
  \begin{subfigure}{0.49\linewidth}
    \includegraphics[width=\linewidth]{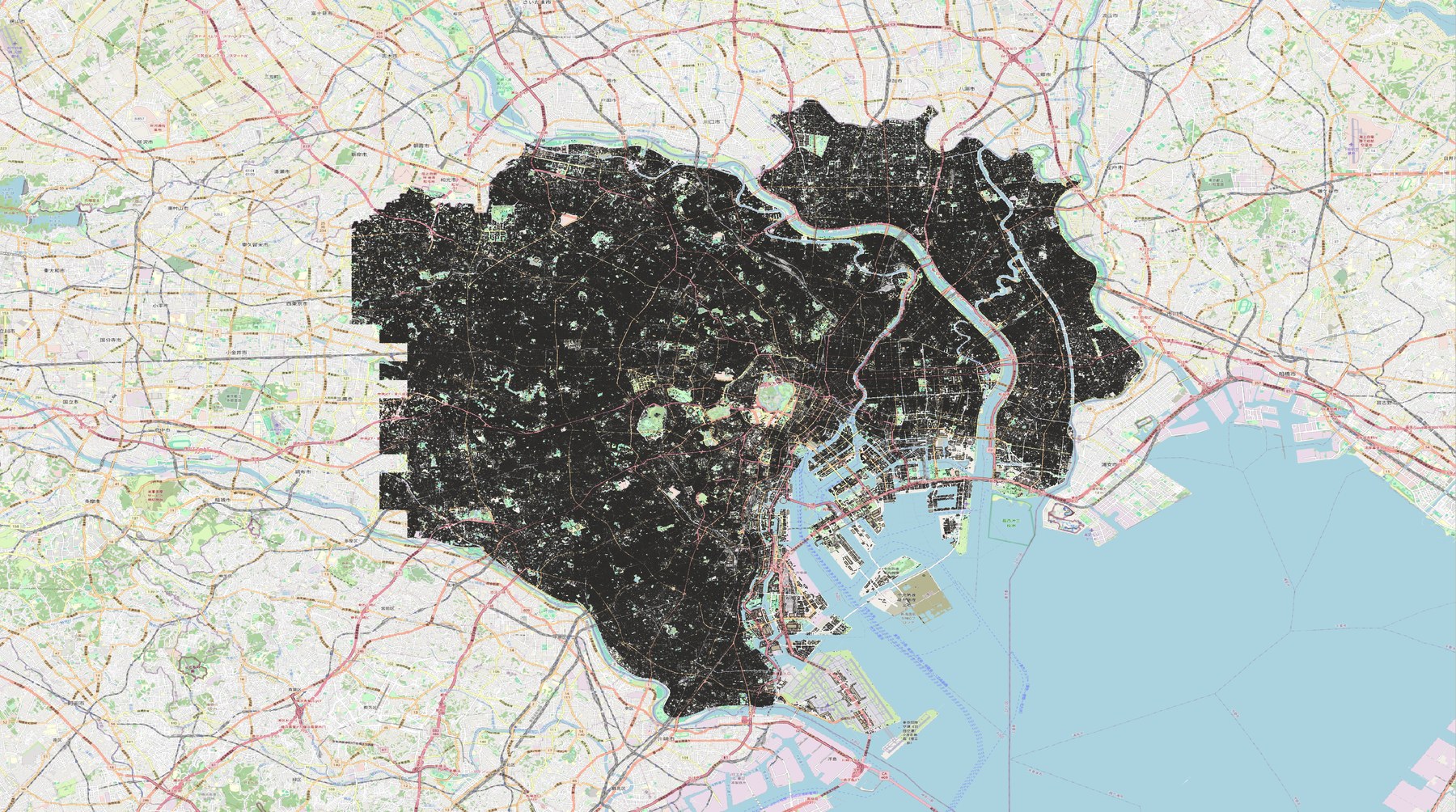}
    \caption{Tokyo (23 wards), 1:46,760}
  \end{subfigure}\\[2pt]
  \begin{subfigure}{0.49\linewidth}
    \includegraphics[width=\linewidth]{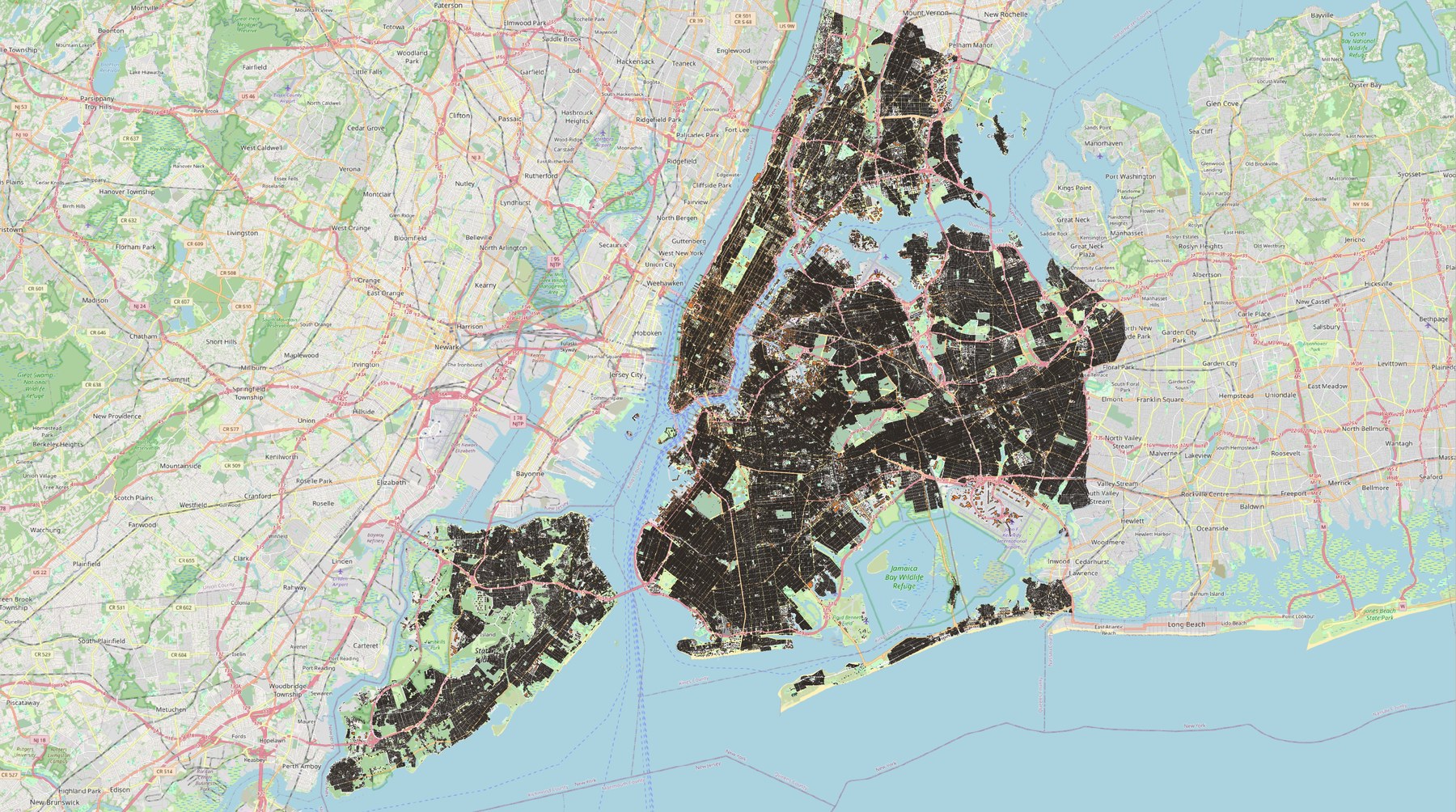}
    \caption{New York, 1:91,455}
  \end{subfigure}\hfill
  \begin{subfigure}{0.49\linewidth}
    \includegraphics[width=\linewidth]{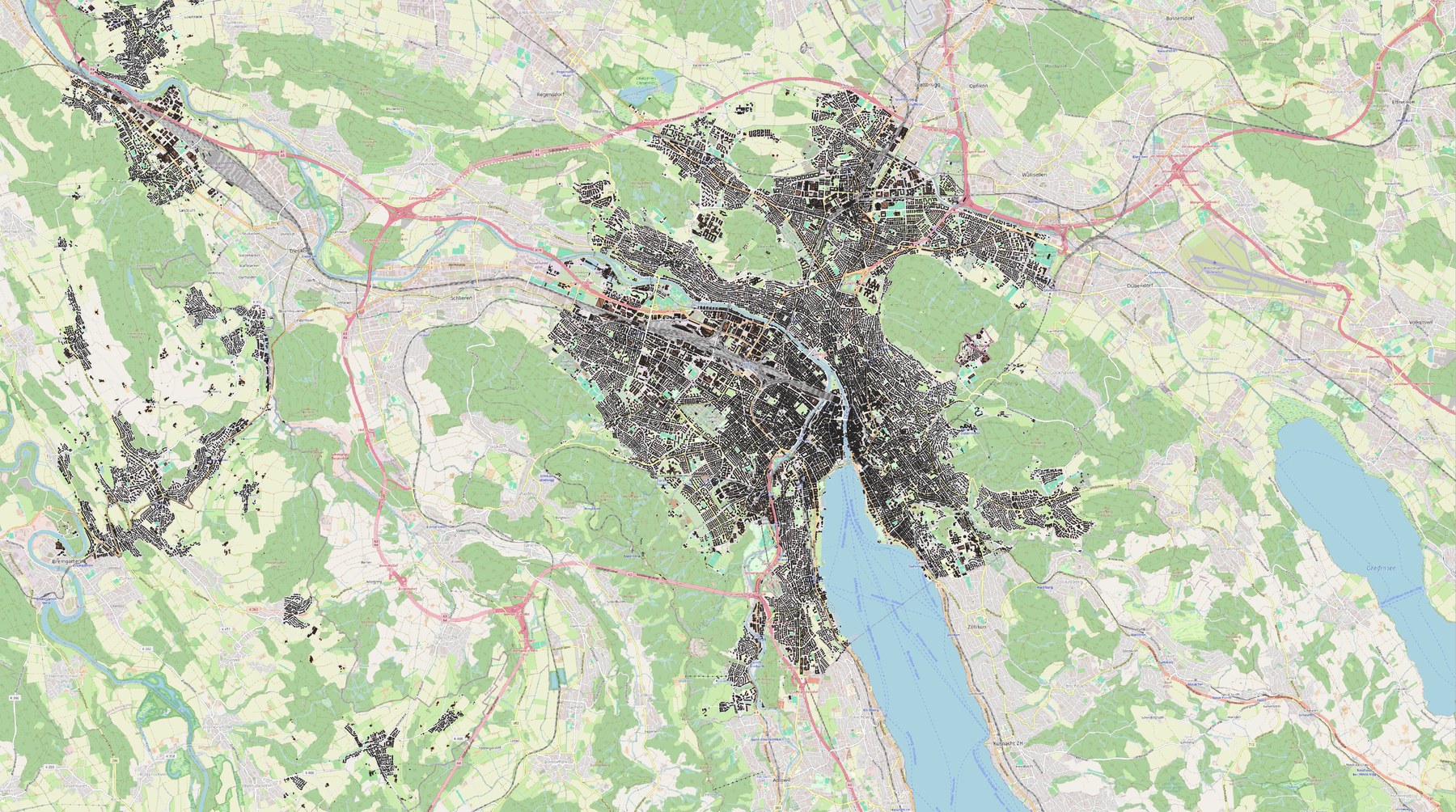} %
    \caption{Zurich, 1:120,480}
  \end{subfigure}\\[2pt]
  \begin{subfigure}{0.49\linewidth}
    \includegraphics[width=\linewidth]{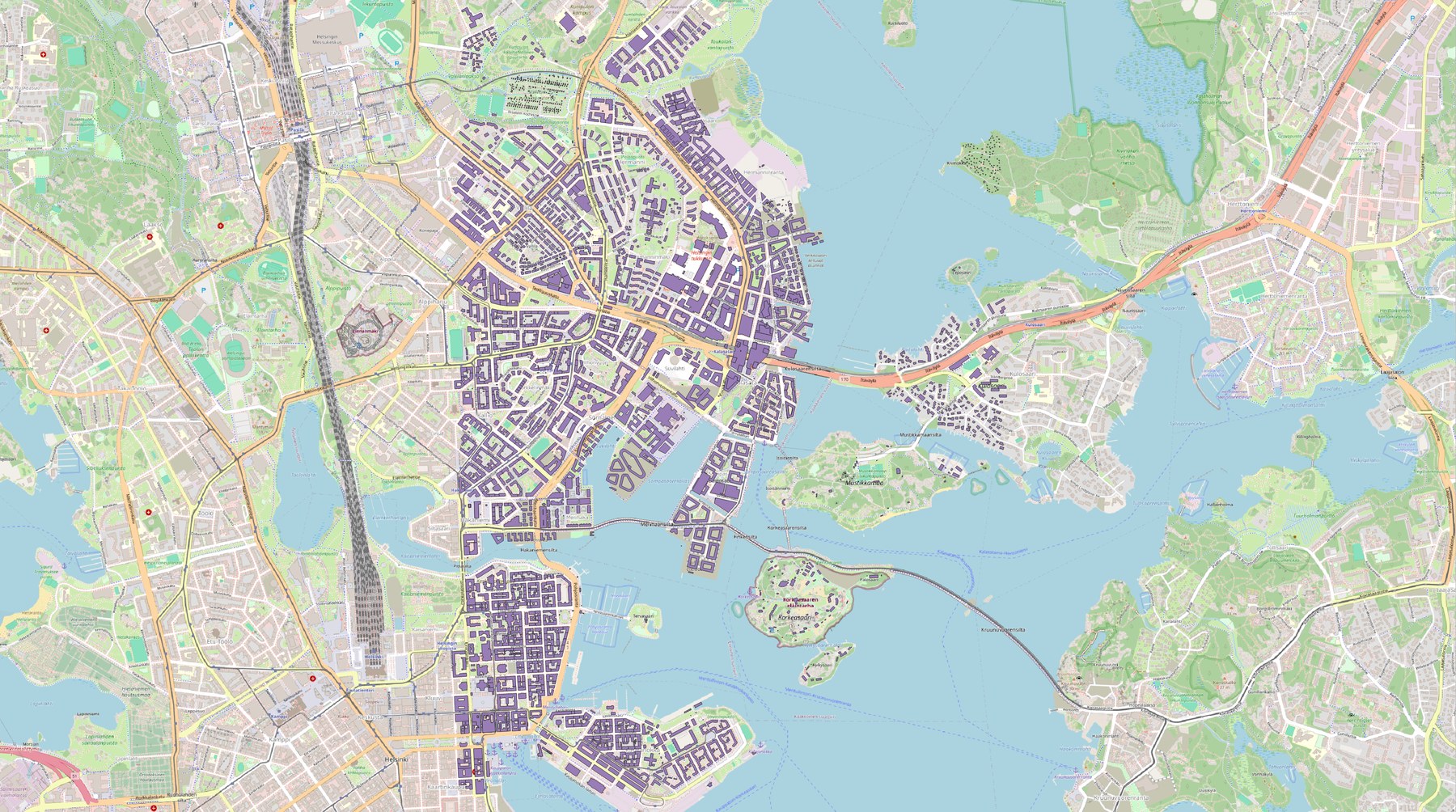}
    \caption{Helsinki (Kalasatama Digital Twins), 1:8734}
  \end{subfigure}
  \caption{
    Spatial extent of the five \benchname{} cities. 
    Basemap: \copyright~OpenStreetMap contributors.
    }
  \Description{Five map panels showing building-footprint extents over OpenStreetMap for Hamburg, Tokyo, New York, Zurich, and Helsinki. The maps use different scales, with Helsinki shown at the most zoomed-in scale and Zurich at the broadest extent.}
  \label{fig:city_extents}
\end{figure}

\section{Datasheet}\label{app:datasheet}

We document \benchname{} following the datasheets framework of~\citet{gebru_datasheets_2021}. Per-city figures referenced below are given in \Cref{tab:city_sources,tab:city_graph_content}. The cities' spatial extents are shown in~\Cref{fig:city_extents}. 

\subsection{Motivation} 

\dsq{For what purpose was the dataset created?} To provide a multi-source, provenance-aware 3D city knowledge graph and a benchmark that evaluates reasoning over source origin, confidence, coverage, and cross-source agreement, which existing text-to-query and urban benchmarks do not exercise (\Cref{sec:related_work}). 

\dsq{Who created it?} \benchname{} was created by the authors of this paper as part of an academic research effort on multi-source urban data integration. 

\subsection{Composition}

\dsq{What do the instances represent?} Nodes are buildings and other city objects (building parts, boundary surfaces, geometry primitives, and, per city, bridges, roads, vegetation, water bodies, and city furniture), together with fused OpenStreetMap features and, for Hamburg, ML-predicted roof materials and reconstructed LoD3 facades.

\dsq{How many instances are there?} Five city graphs totaling roughly \num{180}~GiB, \num{180}M nodes, \num{220}M edges, \num{1.2}B properties, and \num{3.6}M buildings, ranging from \num{2980} buildings (Helsinki) to \num{2005762} (Tokyo).

\dsq{Are relationships between instances made explicit?} Yes, relationships are the dataset. Typed, directed edges connect each building to its parts, boundary surfaces, and geometry primitives, and to its district and city, while cross-source fusion is expressed as explicit \texttt{ENRICHED\_BY}/\texttt{HAS\_POI} edges carrying overlap and confidence attributes rather than as silently merged scalars (\Cref{sec:dataset}). Every edge records its source, so provenance is queryable at the relationship level, not just the node level.

\dsq{Is any information missing, and does the dataset contain all instances or a sample?} It contains all buildings of each released source extent; coverage of derived layers is deliberately partial and flagged (ML roof materials cover \num{50.2}\,\% of Hamburg buildings; LoD3 covers 17 Hamburg buildings), so absence is never a negative label.

\dsq{Does the dataset contain confidential or personal data?} No. Instances are buildings, not people. All sources are already public. Address-level strings are retained only where OpenStreetMap itself publishes them. The remaining datasheet questions concerning human data subjects (consent, ethical review, data retention, offensive content) are therefore not applicable.

\dsq{Are there errors, noise, or redundancies?} Yes, and they are documented and preserved rather than silently altered. Some source datasets ship non-unique \texttt{gml:id}s, resolved at ingest by content and geometry hashing. 
Tokyo's PLATEAU uses a $\pm$\texttt{9999} height sentinel, kept verbatim but excluded from all statistics. The Helsinki source export carries upstream-corrupted Finnish diacritics (\Cref{sec:limitations}).

\subsection{Collection Process}

\dsq{How was the data acquired?} Authoritative CityGML was downloaded from the open-government portals in \Cref{tab:city_sources}. OpenStreetMap was obtained as regional Geofabrik extracts. The roof-material predictions are produced by our own imagery-based model~\cite{lukas_roofmaterials}, and the LoD3 facades are reconstructed from our own facade imagery.

\dsq{Over what timeframe?} The source vintages are listed per city in \Cref{tab:city_sources}. The graph was constructed and fused in 2026.

\subsection{Preprocessing, Cleaning, and Labeling}

\dsq{Was any preprocessing done?} CityGML is mapped to a labeled property graph with coordinates stored verbatim. Non-metric sources (New York, Tokyo) are reprojected to a per-city metric CRS at ingest (\Cref{tab:city_sources}). OpenStreetMap footprints are matched to authoritative footprints by a confidence-weighted overlap graph and attached as \texttt{ENRICHED\_BY} edges without overwriting authoritative values. Derived scalars are prefixed (\texttt{osm\_*}, \texttt{predicted\_*}).

\dsq{Is the raw source available?} Yes, through the sources given in \Cref{tab:city_sources}. The original coordinates, CRS, and \texttt{gml:id}s are retained as provenance so the mapping is auditable.

\dsq{Was the data validated?} Every released instance passes automated integrity gates: a lossless round-trip census on the authoritative layer and an OpenStreetMap ingestion-completeness census on the fused layer (\Cref{subsec:osm_fusion}).

\subsection{Uses}

\dsq{Has the dataset been used for any tasks already?} Yes, the two benchmark families reported in this paper, natural-language-to-query translation (Task~A) and graph representation learning (Task~B), are evaluated on it, and their baseline results are released alongside the dataset (\Cref{subsec:query_tasks,subsec:rl_tasks}). We are aware of no third-party uses at time of release.

\dsq{What tasks is the dataset intended for?} The two benchmark families (\Cref{subsec:query_tasks,subsec:rl_tasks}): natural-language-to-query translation and graph representation learning, both emphasizing provenance-, coverage-, and cross-source-aware reasoning. Additionally, the dataset can serve as a foundation for future graph-based urban analyses and data fusion.

\dsq{What uses should be avoided?} The dataset describes the built environment, not individuals, and should not be repurposed to infer information about residents. Derived layers are marked as predicted or reconstructed precisely so consumers can exclude them where authoritative-only evidence is required.

\subsection{Distribution}

\dsq{How is the dataset distributed and under what license?} As a Neo4j dump (27.2 GiB) plus a backend-neutral node/edge-table export, with loaders, task splits, gold queries and materialized answers, the evaluation harness, and this datasheet (\Cref{sec:dataset}). Each source retains its own license (Hamburg dl-de/by-2-0~\cite{dlde_by20_2026}, Zurich swisstopo~\cite{swisstopo_ogdterms_2022}, Helsinki CC~BY~4.0~\cite{creativecommons_ccby40_2026}, New York NYC OpenData~\cite{nyc_opendata_2026}, Tokyo PLATEAU Public Data License 1.0~\cite{tokyo_license_2024}). OpenStreetMap is licensed under the Open Data Commons Open Database License (ODbL)~\cite{opendatacommons_odbl_2026} by the OpenStreetMap Foundation (OSMF). Our annotations, question suite, and documentation are released under CC~BY~4.0.

\dsq{Is there a DOI?} Yes, the artifact is archived on Zenodo under DOI \href{https://doi.org/10.5281/zenodo.21547211}{\texttt{10.5281/zenodo.21547211}}.

\subsection{Maintenance}

\dsq{Who maintains it and how is it versioned?} The authors host the leaderboard and version the suite. Any change to a gold answer on a dataset rebuild bumps the suite version under a changelog (gate G5, \Cref{app:questions_gate}).

\dsq{Will it be extended?} Yes: further Tier~1 cities and a sensed layer are planned as inference data becomes available (\Cref{sec:limitations}).

\dsq{Contact.} The corresponding author of this paper, Huynh Duc An Son Nguyen (\href{mailto:son.nguyen@hcu-hamburg.de}{son.nguyen@hcu-hamburg.de}).

\section{Limitations}\label{sec:limitations}

\dsq{Geographic scope.} Tier~2 exists for one city and Tier~1 spans five across three continents. Findings may not transfer to regions with different cadastral traditions or OSM community density. 

\dsq{Cross-city heterogeneity.} National schemes differ in detail and vocabulary, so cross-city questions use a common attribute subset (\Cref{sec:dataset}). Per-city attributes remain queryable but outside the comparable core. The binding constraint for the ML-predicted layer is inference-imagery availability. 

\dsq{Upstream source defects.} Source values are preserved verbatim rather than silently repaired: the Helsinki export carries upstream-corrupted Finnish diacritics (\num{1490} Unicode replacement characters across \num{283} values), which we retain as-is while the integrity gates confirm faithful preservation. 

\dsq{Prediction bias.} The roof-material predictions inherit their training-imagery biases~\cite{lukas_roofmaterials} and are marked as predicted. 

\section{CityGML-OSM Knowledge Graph}\label{app:enriched_osm}

\begin{figure*}[h!]
  \centering
  \definecolor{osmFill}{HTML}{57C7E3}\definecolor{osmEdge}{HTML}{23B3D7}%
  \definecolor{cgFill}{HTML}{ECB5C9}\definecolor{cgEdge}{HTML}{DA7298}%
  \definecolor{geoFill}{HTML}{A5ABB6}\definecolor{geoEdge}{HTML}{9AA1AC}%
  \definecolor{poiFill}{HTML}{8DCC93}\definecolor{poiEdge}{HTML}{5DB665}%
  \includegraphics{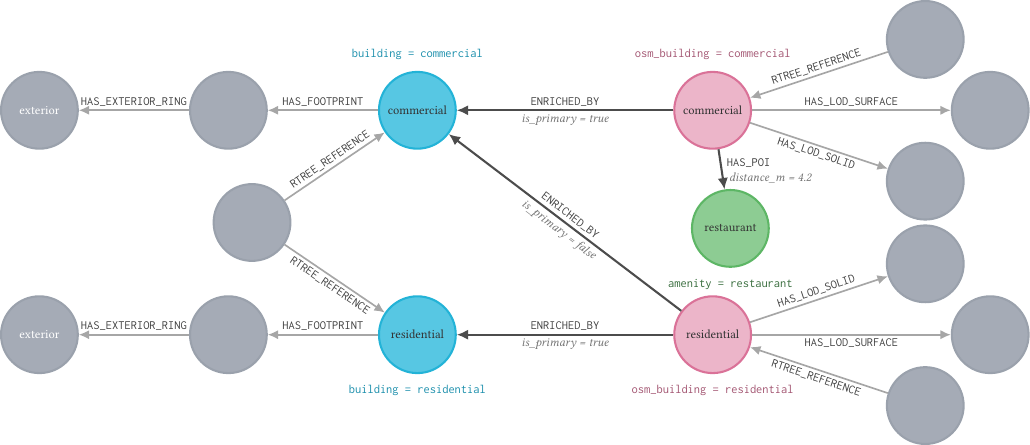}
  \caption{
    OSM enrichment: CityGML buildings (pink) are linked to OSM buildings (blue) by \texttt{ENRICHED\_BY} edges. Buildings inherit source-prefixed \texttt{osm\_*} attributes from their primary match, while points of interest (green) attach via \texttt{HAS\_POI}.
  }
  \Description{Graph diagram with two pink CityGML building nodes on the right connected by ENRICHED_BY edges to two blue OSM building nodes in the center; the lower pink node has an additional non-primary edge to the upper blue node; a green OSM point-of-interest node labeled restaurant hangs below the upper pink node, reached by a HAS_POI edge carrying a distance_m attribute; gray nodes for geometry and spatial index surround both sides.}
  \label{fig:osm_enrich_example}
\end{figure*}

\Cref{fig:osm_enrich_example} illustrates the structure and content of \benchname{}'s LPGs, which extend authoritative CityGML graphs with complementary OSM information. The enrichment process is strictly additive: authoritative CityGML entities and attributes are preserved, while OSM-derived information is attached either as source-prefixed properties on CityGML feature nodes or as separate OSM feature nodes linked through explicit relationships. Building-level correspondences between the two sources are represented through enrichment edges that record spatial matching statistics, including overlap ratio and Jaccard similarity, and identify the primary match based on the largest overlap. To avoid introducing ambiguity, only attributes from the primary OSM match are propagated to the corresponding CityGML building, while information from secondary matches remains associated with its original OSM feature. 
Throughout the graph, provenance is retained at the source level, enabling users to distinguish authoritative CityGML information from OSM-derived enrichments and trace the origin of all integrated data. 

We choose \(\tau\) (\Cref{subsec:osm_fusion}) through a sensitivity analysis of the correspondence structure. Since no city-scale ground-truth CityGML--OSM correspondences exist, we sweep \(\tau \in [0.1, 0.7]\) and evaluate the mean Jaccard of primary matches (precision proxy) and the number of enriched buildings (recall proxy). The mean primary-match Jaccard remains constant (0.775), indicating that \(\tau\) affects only marginal edges. 
We therefore select \(\tau = 0.3\), which provides near-maximal enrichment while preserving correspondence quality. Because each retained edge stores its Jaccard score, downstream methods can reweight or rethreshold matches without repeating the alignment. 

\section{Per-City Dataset and Cross-Source Analysis}\label{app:dataset_stats}

This section gives
an overview comparison of all five datasets~\Cref{fig:city_stats}, 
the full per-city statistics summarized in \Cref{sec:dataset}, 
dataset provenance and scale (\Cref{tab:city_sources}), 
graph size and content (\Cref{tab:city_graph_content}), 
footprint correspondence between CityGML and OSM (\Cref{tab:footprint_cases}), 
and dual-sourced attribute agreement (\Cref{tab:dual_attrs}). 

\begingroup
\definecolor{cham}{HTML}{2A78D6}\definecolor{chel}{HTML}{008300}%
\definecolor{czur}{HTML}{E87BA4}\definecolor{cnyc}{HTML}{EDA100}\definecolor{ctok}{HTML}{1BAF7A}%
\colorlet{cgray}{black!7}%
\tikzset{cbar/.style={rounded corners=0.5pt},
  rl/.style={left,font=\normalsize,inner sep=2.5pt},
  vl/.style={right,font=\small,text=black!55,inner sep=2.5pt}}%
\begin{figure}[h!]
  \centering
  \begin{minipage}[t]{0.5\textwidth}\centering
    {\normalsize\bfseries Detail density: nodes per building}\par\vspace{2pt}
    \includegraphics{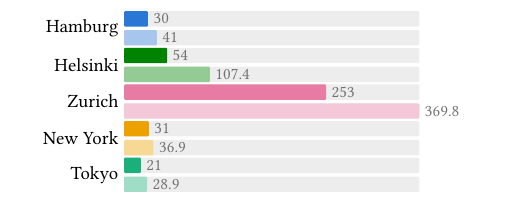}\par\vspace{-2pt}{\small\color{black!55}Darker = median \qquad lighter = mean}\par\vspace{5pt}
  \end{minipage}%
  \hfill 
  \begin{minipage}[t]{0.5\textwidth}\centering
    {\normalsize\bfseries Storage by source layer}\par\vspace{2pt}
    \includegraphics{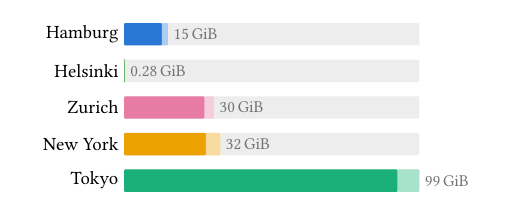}\par\vspace{-2pt}{\small\color{black!55}Darker = CityGML \qquad lighter = $+$ OSM}\par\vspace{5pt}
  \end{minipage}
  \hfill 
  \begin{minipage}[t]{0.5\textwidth}\centering
    {\normalsize\bfseries Source schema richness: thematic keys}\par\vspace{2pt}
    \includegraphics{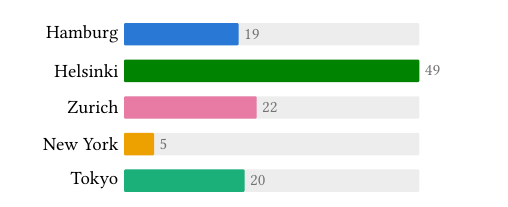}%
  \end{minipage}%
  \hfill 
  \begin{minipage}[t]{0.5\textwidth}\centering
    {\normalsize\bfseries OSM enrichment rate}\par\vspace{2pt}
    \includegraphics{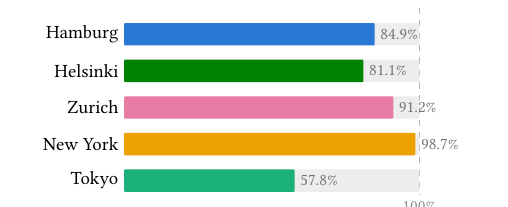}%
  \end{minipage}
  \caption{
    Comparison of the five \benchname{} graphs. Zurich is the densest, Tokyo the largest, Helsinki the most attribute-rich, and New York the most OSM-enriched. Geometry accounts for 85--95\,\% of graph content in every city. 
    }
  \Description{Five side-by-side bar-chart panels comparing Hamburg, Helsinki, Zurich, New York, and Tokyo across graph characteristics: detail density (median and mean nodes per building), schema richness (distinct thematic keys), storage split between CityGML and OSM, OSM enrichment rate percentages, and node composition dominated by geometry.}
  \label{fig:city_stats}
\end{figure}
\endgroup

\providecommand{\citytablefont}{\footnotesize}
\providecommand{\citysourcesfont}{\normalsize}
\begin{table*}[t]
  \caption{Dataset provenance and scale of the five \benchname{} cities. Extent is the bounding box of all building footprints in the source CityGML (WGS84).}
  \label{tab:city_sources}
  \citysourcesfont
  \setlength{\tabcolsep}{5pt}
  \begin{tabular}{l rrrrr r}
    \toprule
    & Hamburg & Helsinki & Zurich & New York & Tokyo & All 5 \\
    & \small DE & \small FI & \small CH & \small US & \small JP & \small corpus \\
    \midrule
    Provider (source) & LGV~\cite{lgv_hhcitygml_2025} & HRI~\cite{helsinki_3dcitymodel_2022} & swisstopo~\cite{swisstopo_swissbuildings3d30_2026} & NYC OTI~\cite{nyc_3dbuildingmodel_2016} & PLATEAU~\cite{plateau_3dcitymodel_2021} & -- \\
    Data vintage & 2025 & 2019 & 2019 & 2016 & 2025 & 2016--2025 \\
    License & dl-de/by-2-0~\cite{dlde_by20_2026} & CC~BY~4.0~\cite{creativecommons_ccby40_2026} & swisstopo~\cite{swisstopo_ogdterms_2022} & NYC OpenData~\cite{nyc_opendata_2026} & PDL 1.0~\cite{tokyo_license_2024} & -- \\
    CityGML / LoD & 2.0 / LoD2 & 2.0 / LoD2 & 2.0 / LoD2.3 & 1.0/2.0 / LoD1--2 & 2.0 / LoD1--3 & -- \\
    Metric CRS & EPSG:25832 & EPSG:3879 & EPSG:2056 & EPSG:32618$^\dagger$ & EPSG:6677$^\dagger$ & -- \\
    Source files (tiles) & \num{1}\textsuperscript{a} & \num{1} & \num{78} & \num{20} & \num{2335} & \num{2435} \\
    Extent (km\textsuperscript{2}, bbox) & \num{7117} & \num{14} & \num{3137} & \num{2159} & \num{1045} & \num{13472} \\
    Source building features & \num{388267} & \num{2980} & \num{102673} & \num{1083437} & \num{2980839} & \num{4558196} \\
    Distinct \texttt{gml:id}s & \num{388267} & \num{2919} & \num{102628} & \num{1083437} & \num{2005294} & \num{3582545} \\
    Building nodes (graph)\textsuperscript{b} & \num{388267} & \num{2980} & \num{102668} & \num{1083437} & \num{2005762} & \num{3583114} \\
    \bottomrule
  \end{tabular}

  \smallskip\raggedright\footnotesize
  \textsuperscript{a} The Hamburg LGV publishes its LoD2 model as map tiles; the single file used here was produced by importing those tiles into \mbox{3DCityDB}~\cite{yao_3dcitydb_2018,yao_3dcitydb5_2025} and re-exporting them as one merged CityGML, so ingest sees a single file whose bounding box is the union of the original tiles. 
  \textsuperscript{b} Nodes representing unique buildings, may differ from total number of input buildings, as in the case of Zurich and Tokyo. 
  $^\dagger$ Reprojected at ingest from a non-metric source CRS (Tokyo EPSG:6697 geographic degrees, New York EPSG:2263 US survey feet). Distinct \texttt{gml:id} counts differ from source feature counts where sources repeat IDs. Identity resolution at ingest splits distinct same-ID features onto synthesized IDs and deduplicates byte-identical replicas (Tokyo's ward-package tiling). Building node counts are thus lossless.
\end{table*}

\providecommand{\citytablefont}{\footnotesize}
\providecommand{\cityenrichfont}{\normalsize}
\begin{table*}[t]
  \caption{Multi-source enrichment of the five city graphs: crowd-source OSM, ML-predicted roof materials, and reconstructed LoD3 geometrical data.}
  \label{tab:city_enrichment}
  \cityenrichfont
  \setlength{\tabcolsep}{4pt}
  \begin{tabular}{l rrrrr r}
    \toprule
    & Hamburg & Helsinki & Zurich & New York & Tokyo & All 5 \\
    & \small DE & \small FI & \small CH & \small US & \small JP & \small corpus \\
    \midrule
    \multicolumn{7}{l}{\textit{Crowd-sourced layer (OSM)}}\\
    OSM features & \num{1188139} & \num{55930} & \num{1860401} & \num{2555468} & \num{3830408} & \num{9490346} \\
    \quad Matched to a city object & 356.7\,k\,\pct{30.0} & \num{4281}\,\pct{7.7} & 102.8\,k\,\pct{5.5} & 1.12\,M\,\pct{43.9} & 1.24\,M\,\pct{32.3} & 2.82\,M\,\pct{29.7} \\
    \quad Standalone net-new\textsuperscript{a} & 831.4\,k\,\pct{70.0} & 51.6\,k\,\pct{92.3} & 1.76\,M\,\pct{94.5} & 1.43\,M\,\pct{56.1} & 2.60\,M\,\pct{67.7} & 6.67\,M\,\pct{70.3} \\
    Buildings enriched (\texttt{ENRICHED\_BY}) & 331.3\,k\,\pct{85.3} & \num{2476}\,\pct{83.1} & 94.0\,k\,\pct{91.5} & 1.07\,M\,\pct{98.8} & 1.18\,M\,\pct{58.9} & 2.68\,M\,\pct{74.8} \\
    \texttt{osm\_*} property records copied & 2.26\,M & 15.4\,k & 601.5\,k & 6.98\,M & 5.12\,M & 14.97\,M \\
    Buildings with $\geq$1 POI & 12.0\,k\,\pct{3.1} & 657\,\pct{22.0} & \num{7569}\,\pct{7.4} & 34.7\,k\,\pct{3.2} & 65.3\,k\,\pct{3.3} & 120.1\,k\,\pct{3.4} \\
    POI attachments (\texttt{HAS\_POI}) & \num{22219} & \num{1797} & \num{13587} & \num{52449} & \num{91487} & \num{181539} \\
    \midrule
    \multicolumn{7}{l}{\textit{ML-predicted layer (roof material; property records on Building nodes)}}\\
    Buildings with prediction & 194.8\,k\,\pct{50.2} & -- & -- & -- & -- & 194.8\,k\,\pct{5.4} \\
    \quad -- thereof multi-material & 570\,\pct{0.3} & -- & -- & -- & -- & -- \\
    Material assignments (ranked) & 195.4\,k & -- & -- & -- & -- & 195.4\,k \\
    Property records written & 975.2\,k & -- & -- & -- & -- & 975.2\,k \\
    \midrule
    \multicolumn{7}{l}{\textit{Reconstructed layer (LoD3)}}\\
    LoD3 facade surfaces & \num{3404} & -- & -- & -- & -- & \num{3404} \\
    Anchor edges (\texttt{HAS\_LOD3\_FACADE}) & \num{303} & -- & -- & -- & -- & \num{303} \\
    \bottomrule
  \end{tabular}

  \smallskip\raggedright\footnotesize
  \textsuperscript{a}\,Standalone nodes are net-new knowledge outside the cadastral layer, kept with an explicit \texttt{unmatched\_reason}: features of a kind that has no CityGML counterpart to anchor to (roads, land use, water, street furniture), point features falling inside no building footprint (\texttt{no\_containing\_building}), and polygons overlapping no CityGML footprint (\texttt{no\_overlap}), e.g.\ Hamburg: \num{752624} / \num{56715} / \num{22093} of its \num{831432} standalone features.
\end{table*}

\providecommand{\citytablefont}{\footnotesize}
\providecommand{\citycontentfont}{\normalsize}
\begin{table*}[p]
  \caption{Graph size and content (nodes and edges) of \benchname{} KGs. Each cell denotes the absolute count and, in parentheses, its share of that city's total nodes or relationships. Predicted roof materials are stored as building properties. } %
  \label{tab:city_graph_content}
  \citycontentfont
  \setlength{\tabcolsep}{3.6pt}
  \definecolor{kgband}{HTML}{DCE3EE}%
  \definecolor{kgcat}{HTML}{EEF2F8}%
  \definecolor{kgagg}{HTML}{EFF4FB}%
  \definecolor{kgSem}{HTML}{2E7D32}%
  \definecolor{kgOsm}{HTML}{1565C0}%
  \definecolor{kgRecon}{HTML}{6A1B9A}%
  \definecolor{kgIdx}{HTML}{6B7280}%
  \begin{tabular}{l rrrrr >{\columncolor{kgagg}}r}
    \toprule
    & Hamburg & Helsinki & Zurich & New York & Tokyo & All 5 \\
    & \small DE & \small FI & \small CH & \small US & \small JP & \small corpus \\
    \midrule
    \rowcolor{kgband}\multicolumn{7}{l}{\textit{Graph size}}\\
    Nodes & \num{17440924} & \num{414459} & \num{41508371} & \num{45797985} & \num{74409400} & \num{179571139} \\
    Relationships & \num{26299391} & \num{619964} & \num{45548405} & \num{54207134} & \num{91949234} & \num{218624128} \\
    Node property records & 113.35\,M & 3.15\,M & 185.79\,M & 291.39\,M & 405.19\,M & 998.88\,M \\
    Relationship property records & 25.60\,M & 513.1\,k & 39.33\,M & 46.18\,M & 73.39\,M & 185.01\,M \\
    \midrule
    \rowcolor{kgband}\multicolumn{7}{l}{\textit{Node content} (count, \% of the city's nodes)}\\
    \rowcolor{kgcat}\textbf{\textcolor{kgSem}{Semantic city objects}} & 388.9\,k\,\pct{2.2} & \num{3050}\,\pct{0.7} & 114.7\,k\,\pct{0.3} & 1.08\,M\,\pct{2.4} & 2.71\,M\,\pct{3.6} & 4.30\,M\,\pct{2.4} \\
    \quad Buildings & 388.3\,k\,\pct{2.2} & \num{2980}\,\pct{0.7} & 102.7\,k\,\pct{0.2} & 1.08\,M\,\pct{2.4} & 2.01\,M\,\pct{2.7} & 3.58\,M\,\pct{2.0} \\
    \quad Building parts & -- & -- & 12.0\,k\,\pct{0.0} & -- & -- & 12.0\,k\,\pct{0.0} \\
    \quad Other city objects\textsuperscript{a} & -- & 31\,\pct{0.0} & -- & -- & 659.0\,k\,\pct{0.9} & 659.0\,k\,\pct{0.4} \\
    \quad Interior \& openings & 499\,\pct{0.0} & -- & -- & -- & 46.2\,k\,\pct{0.1} & 46.7\,k\,\pct{0.0} \\
    \quad Classifier hubs & 138\,\pct{0.0} & 39\,\pct{0.0} & -- & -- & -- & 177\,\pct{0.0} \\
    \rowcolor{kgcat}\textbf{Geometry \& surfaces} & 15.84\,M\,\pct{90.8} & 354.6\,k\,\pct{85.6} & 39.50\,M\,\pct{95.2} & 42.10\,M\,\pct{91.9} & 67.77\,M\,\pct{91.1} & 165.57\,M\,\pct{92.2} \\
    \quad Boundary surfaces & 4.53\,M\,\pct{26.0} & 50.6\,k\,\pct{12.2} & 425.3\,k\,\pct{1.0} & 12.97\,M\,\pct{28.3} & 3.09\,M\,\pct{4.1} & 21.06\,M\,\pct{11.7} \\
    \quad\quad -- roof & 868.0\,k\,\pct{5.0} & 13.4\,k\,\pct{3.2} & 208.4\,k\,\pct{0.5} & 1.58\,M\,\pct{3.5} & 908.0\,k\,\pct{1.2} & 3.58\,M\,\pct{2.0} \\
    \quad\quad -- wall & 3.28\,M\,\pct{18.8} & 34.3\,k\,\pct{8.3} & 108.7\,k\,\pct{0.3} & 10.29\,M\,\pct{22.5} & 1.98\,M\,\pct{2.7} & 15.70\,M\,\pct{8.7} \\
    \quad\quad -- ground & 388.3\,k\,\pct{2.2} & \num{2923}\,\pct{0.7} & 108.2\,k\,\pct{0.3} & 1.09\,M\,\pct{2.4} & 174.4\,k\,\pct{0.2} & 1.77\,M\,\pct{1.0} \\
    \quad\quad -- closure/other & -- & -- & -- & -- & 21.7\,k\,\pct{0.0} & 21.7\,k\,\pct{0.0} \\
    \quad Polygons & 4.95\,M\,\pct{28.4} & 141.1\,k\,\pct{34.0} & 19.25\,M\,\pct{46.4} & 14.34\,M\,\pct{31.3} & 29.56\,M\,\pct{39.7} & 68.25\,M\,\pct{38.0} \\
    \quad Rings & 4.96\,M\,\pct{28.4} & 142.1\,k\,\pct{34.3} & 19.26\,M\,\pct{46.4} & 14.35\,M\,\pct{31.3} & 29.60\,M\,\pct{39.8} & 68.31\,M\,\pct{38.0} \\
    \quad Solids & 388.3\,k\,\pct{2.2} & \num{5585}\,\pct{1.3} & 108.7\,k\,\pct{0.3} & -- & 2.18\,M\,\pct{2.9} & 2.68\,M\,\pct{1.5} \\
    \quad Multi/composite surfaces & 499\,\pct{0.0} & 31\,\pct{0.0} & -- & -- & 2.69\,M\,\pct{3.6} & 2.69\,M\,\pct{1.5} \\
    \quad Line strings & 619.4\,k\,\pct{3.6} & 15.2\,k\,\pct{3.7} & 455.5\,k\,\pct{1.1} & 446.0\,k\,\pct{1.0} & 643.1\,k\,\pct{0.9} & 2.18\,M\,\pct{1.2} \\
    \quad Terrain intersections & 388.2\,k\,\pct{2.2} & -- & -- & -- & -- & 388.2\,k\,\pct{0.2} \\
    \rowcolor{kgcat}\textbf{\textcolor{kgOsm}{Crowd-sourced (OSM features)}} & 1.19\,M\,\pct{6.8} & 55.9\,k\,\pct{13.5} & 1.86\,M\,\pct{4.5} & 2.56\,M\,\pct{5.6} & 3.83\,M\,\pct{5.1} & 9.49\,M\,\pct{5.3} \\
    \rowcolor{kgcat}\textbf{\textcolor{kgRecon}{Reconstructed (LoD3 facades)}} & \num{3404}\,\pct{0.0} & -- & -- & -- & -- & \num{3404}\,\pct{0.0} \\
    \rowcolor{kgcat}\textbf{\textcolor{kgIdx}{Containers \& spatial index}} & 20.4\,k\,\pct{0.1} & 893\,\pct{0.2} & 30.1\,k\,\pct{0.1} & 54.5\,k\,\pct{0.1} & 100.2\,k\,\pct{0.1} & 206.0\,k\,\pct{0.1} \\
    \quad Dataset / City / District & 3\,\pct{0.0} & 3\,\pct{0.0} & 157\,\pct{0.0} & 41\,\pct{0.0} & \num{3201}\,\pct{0.0} & \num{3405}\,\pct{0.0} \\
    \quad R-tree index nodes & 20.4\,k\,\pct{0.1} & 890\,\pct{0.2} & 29.9\,k\,\pct{0.1} & 54.4\,k\,\pct{0.1} & 97.0\,k\,\pct{0.1} & 202.6\,k\,\pct{0.1} \\
    \midrule
    \rowcolor{kgband}\multicolumn{7}{l}{\textit{Thematic (semantic) content\textsuperscript{b}: hub nodes vs.\ edges vs.\ source property records}}\\
    Classifier hubs (function, roof type) & 138 & 39 & -- & -- & -- & 177 \\
    Classifier edges & 776.5\,k & \num{3066} & -- & -- & -- & 779.6\,k \\
    Distinct source thematic keys\textsuperscript{c} & 19 & 49 & 22 & 5 & 20 & -- \\
    Thematic properties on buildings & 6.97\,M & 104.5\,k & 1.94\,M & 5.42\,M & 31.51\,M & 45.94\,M \\
    \midrule
    \rowcolor{kgband}\multicolumn{7}{l}{\textit{Relationship content} (count, \% of the city's relationships)\textsuperscript{d}}\\
    \rowcolor{kgcat}\textbf{Geometry composition} & 20.37\,M\,\pct{77.5} & 434.4\,k\,\pct{70.1} & 39.50\,M\,\pct{86.7} & 42.10\,M\,\pct{77.7} & 71.50\,M\,\pct{77.8} & 173.92\,M\,\pct{79.6} \\
    \rowcolor{kgcat}\textbf{\textcolor{kgSem}{Semantic / thematic links}} & 777.0\,k\,\pct{3.0} & \num{3066}\,\pct{0.5} & 12.0\,k\,\pct{0.0} & -- & 46.2\,k\,\pct{0.1} & 838.4\,k\,\pct{0.4} \\
    \rowcolor{kgcat}\textbf{\textcolor{kgIdx}{Topology, containment \& index}} & 4.75\,M\,\pct{18.1} & 177.7\,k\,\pct{28.7} & 5.92\,M\,\pct{13.0} & 10.97\,M\,\pct{20.2} & 18.92\,M\,\pct{20.6} & 40.74\,M\,\pct{18.6} \\
    \rowcolor{kgcat}\textbf{\textcolor{kgOsm}{Multi-source fusion}} & 395.7\,k\,\pct{1.5} & \num{4808}\,\pct{0.8} & 113.9\,k\,\pct{0.3} & 1.13\,M\,\pct{2.1} & 1.48\,M\,\pct{1.6} & 3.13\,M\,\pct{1.4} \\
    \midrule
    \rowcolor{kgband}\multicolumn{7}{l}{\textit{Per-building detail} (median\textsuperscript{e})}\\
    Nodes / building & 30 & 54 & 253 & 31 & 21 & -- \\
    Geometry polygons / building & 8 & 20 & 124 & 10 & 9 & -- \\
    Source thematic attributes / building & 18 & 41 & 20 & 5 & 16 & -- \\
    Measured height\textsuperscript{f} (m) & 7.53 & 10.11 & 9.22 & -- & 7.9 & -- \\
    \bottomrule
  \end{tabular}

  \smallskip\raggedright\footnotesize
  \textsuperscript{a}\,Helsinki: \num{31} bridges. Tokyo (B-core module set): \num{605293} roads, \num{38231} city furniture, \num{10883} vegetation objects, \num{968} bridges, \num{2895} water bodies, \num{735} plant cover, \num{1} city object group.\\
  \textsuperscript{b}\,Excluding IDs, bounding boxes, spatial indices, and enriched properties (OSM, predicted roof materials, LoD3). 
  \textsuperscript{c}\,Property keys that originate in the input CityGML (core attributes and generic attributes), which may carry ML/DL-relevant semantics; Every key is inventoried per city in \Cref{tab:city_thematic_attrs}. 
  \textsuperscript{d}\,\emph{Geometry composition}: \texttt{HAS\_BOUNDARY}, \texttt{HAS\_POLYGON}, \texttt{HAS\_SURFACE\_MEMBER}, \texttt{HAS\_EXTERIOR\_/INTERIOR\_RING}, \texttt{HAS\_LOD\_SOLID/SURFACE/GEOMETRY}, \texttt{HAS\_FOOTPRINT}, \texttt{HAS\_LINE}, \texttt{HAS\_TERRAIN\_INTERSECTION}. \emph{Semantic/thematic}: \texttt{HAS\_FUNCTION}, \texttt{HAS\_ROOF\_TYPE}, \texttt{HAS\_BUILDING\_PART}, \texttt{HAS\_OPENING}, installation/room/nesting edges. \emph{Topology, containment, and index}: \texttt{PART\_OF}, \texttt{LOCATED\_IN}, \texttt{COVERS}, \texttt{RTREE\_*} (spatial R-tree). \emph{Multi-source fusion}: \texttt{ENRICHED\_BY}, \texttt{HAS\_POI}, \texttt{HAS\_LOD3\_FACADE}. 
  \textsuperscript{e}\,Per-building figures are medians (robust to a few hub \texttt{Building} nodes with heavy payloads). 
  \textsuperscript{f}\,Per-building \texttt{bldg:measuredHeight} attribute (height above ground per the national definition, not terrain elevation). 
\end{table*}

\begin{table}[h!]
  \caption{
  Footprint correspondence between CityGML and OSM buildings.
  }
  \label{tab:footprint_cases}
  \begin{tabular}{lrrrr}
    \toprule
    Case & Components & Share & CityGML bldgs & OSM bldgs \\
    \midrule
    \multicolumn{5}{l}{\emph{Hamburg}} \\
    1:1 & \num{256219} & 87.2\,\% & \num{255984} & \num{255984} \\
    1:n & \num{19167} & 6.5\,\% & \num{47936} & \num{19096} \\
    n:1 & \num{16146} & 5.5\,\% & \num{16142} & \num{47838} \\
    n:m & \num{2363} & 0.8\,\% & \num{9537} & \num{9151} \\
    \midrule
    \multicolumn{5}{l}{\emph{Helsinki}} \\
    1:1 & \num{1692} & 84.2\,\% & \num{1686} & \num{1686} \\
    1:n & \num{127} & 6.3\,\% & \num{285} & \num{125} \\
    n:1 & \num{129} & 6.4\,\% & \num{127} & \num{281} \\
    n:m & \num{62} & 3.1\,\% & \num{318} & \num{302} \\
    \midrule
    \multicolumn{5}{l}{\emph{Zurich}} \\
    1:1 & \num{75541} & 90.8\,\% & \num{75430} & \num{75430} \\
    1:n & \num{4858} & 5.8\,\% & \num{12442} & \num{4853} \\
    n:1 & \num{2003} & 2.4\,\% & \num{1999} & \num{4325} \\
    n:m & \num{787} & 0.9\,\% & \num{3812} & \num{3744} \\
    \midrule
    \multicolumn{5}{l}{\emph{New York}} \\
    1:1 & \num{1053850} & 99.3\,\% & \num{1053727} & \num{1053727} \\
    1:n & \num{3046} & 0.3\,\% & \num{8448} & \num{3015} \\
    n:1 & \num{3199} & 0.3\,\% & \num{3178} & \num{6149} \\
    n:m & \num{850} & 0.1\,\% & \num{4277} & \num{4351} \\
    \midrule
    \multicolumn{5}{l}{\emph{Tokyo}} \\
    1:1 & \num{823560} & 85.9\,\% & \num{821655} & \num{821655} \\
    1:n & \num{41392} & 4.3\,\% & \num{90648} & \num{40561} \\
    n:1 & \num{22288} & 2.3\,\% & \num{22273} & \num{46975} \\
    n:m & \num{71562} & 7.5\,\% & \num{224653} & \num{224226} \\
    \bottomrule
  \end{tabular}

  \smallskip\raggedright\footnotesize
  Reading: in 1:n one OSM polygon covers several cadastral buildings; in n:1 several OSM polygons tile one cadastral building. The case mix is itself a per-city property spanning more than an order of magnitude: the fragmented share (n:1 plus n:m) is 0.4\,\% for New York (whose crowd mapping tracks the cadastre almost one-to-one), 3.3\,\% for Zurich, 6.3\,\% for Hamburg, 9.5\,\% for Helsinki, and 9.8\,\% in Tokyo, whose crowd mapping splits more cadastral buildings into several tagged parts.
\end{table}

\begin{table}[h!]
  \caption{
  Dual-sourced building attributes when a CityGML, OSM or ML-predicted value describe the same property. 
  }
  \label{tab:dual_attrs}
  \begin{tabularx}{\linewidth}{llrX}
    \toprule
    Property & City & Dual & Agreement or $|\Delta|$\\
    \midrule
    Storeys & Hamburg & \num{144364} & 87.2\,\% exact; 98.6\,\% $\pm$1 \\
    & Helsinki & \num{48} & 70.8\,\% exact; 91.7\,\% $\pm$1 \\
    & Tokyo & \num{116022} & 85.9\,\% exact; 94.3\,\% $\pm$1 \\
    \addlinespace
    Height & Hamburg & \num{3404} & 85.0\,\% $\leq$ 2\,m; 6.4\,\% $>$ 5\,m \\
    & Helsinki & \num{76} & 61.8\,\% $\leq$ 2\,m; 28.9\,\% $>$ 5\,m \\
    & Zurich & \num{770} & 46.4\,\% $\leq$ 2\,m; 22.6\,\% $>$ 5\,m \\
    & Tokyo & \num{61373} & 94.4\,\% $\leq$ 2\,m; 2.5\,\% $>$ 5\,m \\
    \addlinespace
    Roof mat. & Hamburg & \num{3641} & 75.2\,\% over 5 classes \\
    \bottomrule
  \end{tabularx}
\end{table}

\section{Source Thematic Property Inventory}\label{app:thematic}

Every property key that originates in each city's source CityGML (core attributes plus \texttt{gen:*} generic attributes), with the number of buildings carrying it, is inventoried in \Cref{tab:city_thematic_attrs}. These verbatim, multilingual source vocabularies (German ALKIS/AdV codes, Finnish, Swiss-German, a sparse English LoD1--2 schema, and Japanese PLATEAU keys) are the semantics a text-to-query model must bridge (\Cref{subsec:query_tasks}) and the attributes available for representation-learning tasks (\Cref{subsec:rl_tasks}).

\providecommand{\citytablefont}{\footnotesize}
\providecommand{\citythematicfont}{\normalsize}
\begin{table*}[t]
  \caption{
  Overview of every thematic property key (core and generic attributes) originating from input CityGML datasets with the number of buildings carrying it. 
  Keys are verbatim source vocabulary in each portal's own language.
  }
  \label{tab:city_thematic_attrs}
  \citythematicfont
  \setlength{\tabcolsep}{6pt}\renewcommand{\arraystretch}{1.25}
  \begin{tabularx}{\linewidth}{@{}l >{\raggedright\arraybackslash}X@{}}
    \toprule
    City & Source thematic property keys \footnotesize(buildings carrying the key) \\
    \midrule
    \textbf{Hamburg}\,\footnotesize(19) & \mbox{\texttt{address\_xml}\,388.3\,k}, \mbox{\texttt{creation\_date}\,388.3\,k}, \mbox{\texttt{datenquellebodenhoehe}\,388.3\,k}, \mbox{\texttt{datenquelledachhoehe}\,388.3\,k}, \mbox{\texttt{datenquellelage}\,388.3\,k}, \mbox{\texttt{description}\,388.3\,k}, \mbox{\texttt{external\_information\_system}\,388.3\,k}, \mbox{\texttt{external\_object\_name}\,388.3\,k}, \mbox{\texttt{external\_reference\_xml}\,388.3\,k}, \mbox{\texttt{function\_code}\,388.3\,k}, \mbox{\texttt{gemeindeschluessel}\,388.3\,k}, \mbox{\texttt{geometrietyp2dreferenz}\,388.3\,k}, \mbox{\texttt{grundrissaktualitaet}\,388.3\,k}, \mbox{\texttt{measured\_height}\,388.3\,k}, \mbox{\texttt{measured\_height\_uom}\,388.3\,k}, \mbox{\texttt{roof\_type\_code}\,388.3\,k}, \mbox{\texttt{datenquellegeschossanzahl}\,375.7\,k}, \mbox{\texttt{storeys\_above\_ground}\,375.7\,k}, \mbox{\texttt{gml\_name}\,\num{1665}} \\
    \addlinespace
    \textbf{Helsinki}\,\footnotesize(49) & \mbox{\texttt{creation\_date}\,\num{2980}}, \mbox{\texttt{description}\,\num{2980}}, \mbox{\texttt{address\_xml}\,\num{2972}}, \mbox{\texttt{external\_information\_system}\,\num{2972}}, \mbox{\texttt{external\_object\_name}\,\num{2972}}, \mbox{\texttt{external\_reference\_xml}\,\num{2972}}, \mbox{\texttt{groundlevel}\,\num{2972}}, \mbox{\texttt{highestroof}\,\num{2972}}, \mbox{\texttt{lowestroof}\,\num{2972}}, \mbox{\texttt{measured\_height}\,\num{2972}}, \mbox{\texttt{measured\_height\_uom}\,\num{2972}}, \mbox{\texttt{roof\_type\_code}\,\num{2972}}, \mbox{\texttt{area\_diff}\,\num{2846}}, \mbox{\texttt{area\_diff\_filter}\,\num{2846}}, \mbox{\texttt{file\_candidate\_gmlid}\,\num{2846}}, \mbox{\texttt{integrating\_person}\,\num{2846}}, \mbox{\texttt{integration\_date}\,\num{2846}}, \mbox{\texttt{matching\_mode}\,\num{2846}}, \mbox{\texttt{overlap\_db\_to\_file}\,\num{2846}}, \mbox{\texttt{overlap\_file\_to\_db}\,\num{2846}}, \mbox{\texttt{overlap\_filter}\,\num{2846}}, \mbox{\texttt{code}\,\num{2727}}, \mbox{\texttt{repaired}\,\num{2727}}, \mbox{\texttt{uuid}\,\num{2727}}, \mbox{\texttt{brec\_buildingheightnn}\,\num{2605}}, \mbox{\texttt{brec\_roofnames}\,\num{2605}}, \mbox{\texttt{gen\_id}\,\num{2176}}, \mbox{\texttt{rakennuksen\_tila}\,\num{2176}}, \mbox{\texttt{c\_kayttark}\,\num{1931}}, \mbox{\texttt{kerroksia}\,\num{1931}}, \mbox{\texttt{kg\_krakenn}\,\num{1931}}, \mbox{\texttt{rakennustunnus\_\_ratu}\,\num{1931}}, \mbox{\texttt{rakennustunnus\_\_vtj\_prt}\,\num{1931}}, \mbox{\texttt{tila\_koodi}\,\num{1931}}, \mbox{\texttt{kayttotarkoitus}\,\num{1809}}, \mbox{\texttt{kerrosala}\,\num{1809}}, \mbox{\texttt{valmistunut}\,\num{1752}}, \mbox{\texttt{katuosoite}\,\num{1750}}, \mbox{\texttt{tilavuus}\,\num{1743}}, \mbox{\texttt{kokonaisala}\,\num{1732}}, \mbox{\texttt{rakennusaine}\,\num{1729}}, \mbox{\texttt{buildingheightnn}\,367}, \mbox{\texttt{roofnames}\,367}, \mbox{\texttt{suunnitelma\_alue}\,245}, \mbox{\texttt{kerrosala\_\_m2}\,122}, \mbox{\texttt{tila}\,122}, \mbox{\texttt{tyyppi}\,122}, \mbox{\texttt{function\_code}\,94}, \mbox{\texttt{storeys\_above\_ground}\,92} \\
    \addlinespace
    \textbf{Zurich}\,\footnotesize(22) & \mbox{\texttt{description}\,102.7\,k}, \mbox{\texttt{egid}\,99.3\,k}, \mbox{\texttt{dach\_max}\,96.7\,k}, \mbox{\texttt{dach\_min}\,96.7\,k}, \mbox{\texttt{datum\_aenderung}\,96.7\,k}, \mbox{\texttt{datum\_erstellung}\,96.7\,k}, \mbox{\texttt{erstellung\_jahr}\,96.7\,k}, \mbox{\texttt{erstellung\_monat}\,96.7\,k}, \mbox{\texttt{gen\_predicate}\,96.7\,k}, \mbox{\texttt{grund\_aenderung}\,96.7\,k}, \mbox{\texttt{herkunft}\,96.7\,k}, \mbox{\texttt{herkunft\_jahr}\,96.7\,k}, \mbox{\texttt{herkunft\_monat}\,96.7\,k}, \mbox{\texttt{objektart}\,96.7\,k}, \mbox{\texttt{original\_herkunft}\,96.7\,k}, \mbox{\texttt{revision\_jahr}\,96.7\,k}, \mbox{\texttt{revision\_monat}\,96.7\,k}, \mbox{\texttt{gelaendepunkt}\,96.7\,k}, \mbox{\texttt{measured\_height}\,96.0\,k}, \mbox{\texttt{measured\_height\_uom}\,96.0\,k}, \mbox{\texttt{usage\_code}\,221}, \mbox{\texttt{gml\_name}\,125} \\
    \addlinespace
    \textbf{New York}\,\footnotesize(5) & \mbox{\texttt{bin}\,1.08\,M}, \mbox{\texttt{description}\,1.08\,M}, \mbox{\texttt{doitt\_id}\,1.08\,M}, \mbox{\texttt{gml\_name}\,1.08\,M}, \mbox{\texttt{source\_id}\,1.08\,M} \\
    \addlinespace
    \textbf{Tokyo}\,\footnotesize(20) & \mbox{\texttt{ade\_xml}\,2.01\,M}, \mbox{\texttt{class\_code}\,2.01\,M}, \mbox{\texttt{class\_codespace}\,2.01\,M}, \mbox{\texttt{creation\_date}\,2.01\,M}, \mbox{\texttt{description}\,2.01\,M}, \mbox{\texttt{measured\_height}\,2.01\,M}, \mbox{\texttt{measured\_height\_uom}\,2.01\,M}, \mbox{\texttt{storeys\_above\_ground}\,2.01\,M}, \mbox{\texttt{storeys\_below\_ground}\,2.01\,M}, \mbox{\texttt{usage\_code}\,2.01\,M}, \mbox{\texttt{usage\_codespace}\,2.01\,M}, \mbox{\texttt{address\_xml}\,1.81\,M}, \jp{13\_区市町村コード\_大字\_町コード\_町\_丁目コード}\,(municipality–oaza–chome code)$^\dagger$\,1.81\,M, \jp{大字\_町コード}\,(oaza–town code)$^\dagger$\,1.81\,M, \jp{延べ面積換算係数}\,(total floor-area conversion factor)$^\dagger$\,1.78\,M, \jp{町\_丁目コード}\,(town–chome code)$^\dagger$\,1.72\,M, \jp{地区計画}\,(district plan)$^\dagger$\,324.1\,k, \jp{説明注記}\,(descriptive note)$^\dagger$\,184.1\,k, \jp{再開発等促進区を定める地区計画}\,(redevelopment-promotion district plan)$^\dagger$\,\num{6938}, \mbox{\texttt{gml\_name}\,\num{4909}} \\
    \addlinespace
    \bottomrule
  \end{tabularx}

  \smallskip\raggedright\footnotesize
  $^\dagger$\,The parenthetical English is our translation. The Japanese key is stored verbatim in the graph and is what a query must match.
\end{table*}

\section{Question Suite Design and Statistics}\label{app:questions}

This section summarizes the question suite specification. The full document (all 84 template definitions with gold-query sketches, slot providers, sizing model, and freeze process) ships with the benchmark repository. \Cref{tab:categories}'s instance counts are the frozen, live-verified v1.0 numbers.

\begin{table*}[h!]
  \caption{
    Question categories with template and instance counts, aggregated across all five cities.  
    }
  \label{tab:categories}
  \begin{tabularx}{\textwidth}{lXcrr}
    \toprule
    Category & Capability probed & Tier & Tmpl. & Instances \\
    \midrule
    Aggregate / filter / top-k & schema grounding, retrieval, ranking, incl.\ native per-city attributes 
              & 1 & 18 & 230 \\
    Multi-hop semantic & traversal over the compact schema, incl.\ native per-city attributes 
              & 1 & 11 & 105 \\
    Spatial & window, proximity, geometric multi-hop, 3D structure, density & 1 & 12 & 462 \\
    Cross-source & (dis)agreement, coverage gaps, address completeness, dual roof shapes, OSM-only
              & 1 & 13 & 237 \\
    Provenance-filtered & source and confidence constraints, provenance metadata
            & 1 & 12 & 162 \\
    Coverage-aware & partial-coverage reasoning, incl.\ multi-material predictions & 2 & 6 & 24 \\
    LoD3 showcase & facade/interior structure & 2 & 4 & 21 \\
    Infeasible & hallucination resistance (guard-verified) & 1 & 8 & 153 \\
    Multilingual & German/Japanese language variants; Japanese-key gold queries & 1 & 3 & 95 \\
    \midrule
    Total & & & 84 & \num{1394} \\
    \bottomrule
  \end{tabularx}

  \smallskip\raggedright\footnotesize
  Tier~2 categories are available only for Hamburg. \emph{Multilingual} is an annotation applied across existing templates (German, Japanese, etc.) and is therefore excluded from the total.
\end{table*}

\subsection{Template Inventory}\label{app:questions_inventory}

The categories of \Cref{tab:categories} are divided into capability subfamilies. The five distinctive categories are deliberately the deepest: spatial alone accounts for roughly a third of the feasible instances, and the five together for approximately 70\,\%. Templates marked cross-city use only the common attribute subset of \Cref{sec:dataset} and instantiate on all five cities. City-specific templates (ALKIS code lists, Japanese-keyed attributes, prediction and LoD3 layers) are tagged as such and instantiate only where their layers exist, a constraint the generator asserts at run time.

\subsection{Difficulty Rubric}\label{app:questions_difficulty}

Each question is assigned an authored difficulty level (\Cref{tab:app_difficulty}). 
After completing the baseline evaluation matrix, we additionally report the observed difficulty of each question, measured by its failure rate across models, and quantify the correlation between authored and observed difficulty. Any discrepancy between the two is treated as an empirical finding rather than a reason to revise the original difficulty labels.
Gold answers exhibit a diverse set of result formats (\Cref{tab:app_answer_types}), determined by the structure of the corresponding gold query outputs. Single-value results are the most common, including aggregates, counts, and scalar lookups. These are followed by multi-column or grouped tables, ranked top-$k$ lists, and infeasible cases that require refusal. Notably, none of the benchmark templates produces a bare boolean answer.

\begin{table}
  \caption{Authored difficulty rubric and its distribution over the \num{1394} frozen question instances. Conditions of lower levels may also apply at higher ones.}
  \label{tab:app_difficulty}
  \begin{tabularx}{\linewidth}{cXr}
    \toprule
    Level & Definition & Questions \\
    \midrule
    1 & one node label, one property, one aggregate; no filter & 47 (3.4\%) \\
    2 & one filter, one relationship hop, or ranking (\texttt{ORDER BY}/\texttt{LIMIT}) & 272 (19.5\%) \\
    3 & two composed elements: two hops, a grouped aggregate, one spatial predicate, one edge-property (provenance) filter, or a missing-value subtlety & 714 (51.2\%) \\
    4 & cross-concern composition: spatial $\times$ semantic, cross-source comparison with computation, coverage normalization, percentile or conditional share, schema bridging across languages & 247 (17.7\%) \\
    5 & three or more schema regions plus a geometric or set-level predicate: geometric multi-hop joins, per-entity grouped top-1 over two-hop patterns, n:m correspondence reasoning & 114 (8.2\%) \\
    \bottomrule
  \end{tabularx}
\end{table}

\begin{table}
  \caption{Gold answer-type distribution over the 1394 frozen question instances.}
  \label{tab:app_answer_types}
  \begin{tabularx}{\linewidth}{Xr}
    \toprule
    Answer type & Questions \\
    \midrule
    Scalar, numeric & 556 (39.9\%) \\
    Table (multi-column or grouped) & 352 (25.3\%) \\
    Ranked list (top-$k$, ordered and limited) & 259 (18.6\%) \\
    Refusal (infeasible) & 153 (11.0\%) \\
    Scalar, string & 48 (3.4\%) \\
    List (single column) & 26 (1.9\%) \\
    \bottomrule
  \end{tabularx}
\end{table}

\subsection{Per-City Feasibility}\label{app:questions_matrix}

\Cref{tab:app_matrix} specifies which benchmark categories are instantiated for each city. Missing entries are intentional design choices.
In particular, the coverage and LoD3 rows define the Tier~2 deep-fusion categories. Likewise, the absent-layer infeasible subfamily relies on the absence of prediction layers outside Hamburg, allowing the same question template to be feasible in one city and infeasible in another while keeping the question text unchanged.

\begin{table}
  \caption{Category $\times$ city feasibility. HH Hamburg, ZH Zurich, HEL Helsinki, TYO Tokyo, NYC New York.}
  \label{tab:app_matrix}
  \begin{tabular}{lccccc}
    \toprule
    Category & HH & ZH & HEL & TYO & NYC \\
    \midrule
    Aggregate / filter / multi-hop & \CIRCLE & \CIRCLE & \CIRCLE & \CIRCLE & \CIRCLE \\
    Spatial\textsuperscript{1} & \CIRCLE & \CIRCLE & \CIRCLE & \CIRCLE & \CIRCLE \\
    Cross-source / provenance\textsuperscript{2} & \CIRCLE & \CIRCLE & \CIRCLE & \CIRCLE & \CIRCLE \\
    Coverage-aware & \CIRCLE & -- & -- & -- & -- \\
    LoD3 showcase\textsuperscript{3} & \CIRCLE & -- & -- & -- & -- \\
    Infeasible (other subfamilies) & \CIRCLE & \CIRCLE & \CIRCLE & \CIRCLE & \CIRCLE \\
    Infeasible (absent layer) & -- & \CIRCLE & \CIRCLE & \CIRCLE & \CIRCLE \\
    Multilingual schema bridging & -- & -- & -- & \CIRCLE & -- \\
    Dual roof-shape reporting & \CIRCLE & \CIRCLE & -- & -- & -- \\
    LoD3 confidence scalars\textsuperscript{3} & \CIRCLE & -- & -- & -- & -- \\
    Native single-city content\textsuperscript{4} & -- & \CIRCLE & \CIRCLE & \CIRCLE & -- \\
    OSM-only height\textsuperscript{5} & \CIRCLE & \CIRCLE & \CIRCLE & \CIRCLE & \CIRCLE \\
    \bottomrule
  \end{tabular}
  \par\smallskip\raggedright\footnotesize
  \textsuperscript{1} After metric-CRS reprojection at ingest for Tokyo and New York (\Cref{subsec:construction}). 
  \textsuperscript{2} After whole-city OSM fusion. 
  \textsuperscript{3} Applicable only to selected buildings in Hamburg (\Cref{subsec:predicted_layers}). 
  \textsuperscript{4} Single-home-city templates, gated by \Cref{subsec:construction}'s requirement gate. Helsinki: floor area, storeys, volume, material, integration method, building status; Zurich: data vintage, revision history; Tokyo: basement storeys, town code, redevelopment district. Hamburg's own single-city content is the coverage and LoD3 rows above. 
  \textsuperscript{5} Cross-city by construction (no per-city gate needed), but only substantive where a real gap exists between the two sources' coverage, most pronounced for New York. 
\end{table}

\subsection{Verification Gate}\label{app:questions_gate}

Every question passes five machine-checked validation gates before entering the benchmark suite. \textbf{G1} verifies that the gold query executes successfully without error. \textbf{G2} requires the canonicalized answer hash to be identical across three independent executions, ensuring result stability. \textbf{G3} checks that the result is non-empty unless the underlying template explicitly permits empty outputs. \textbf{G4}, applied to infeasible questions, executes a guard query that must demonstrate the absence of the requested information in the target city's graph by returning a count of zero. \textbf{G5} enforces benchmark versioning: whenever the dataset is rebuilt, all gold answers are re-materialized, and any change triggers a patch-version increment together with a changelog entry, preventing gold answers from silently drifting from the released graph. In addition to these automated checks, we perform human validation on a stratified sample comprising at least 20\% of questions, sampled across categories, difficulty levels, and cities. Each sampled question is independently reviewed 
for gold-answer correctness and natural language quality.
At benchmark freeze time, we document the rejection rates for each validation gate together with the human-review acceptance rates.

\subsection{Example Questions with Gold Queries}\label{app:questions_examples}

One representative example is provided for each distinctive capability. Placeholders enclosed in angle brackets are instantiated at generation time using values retrieved directly from the released graph. 

\emph{Spatial (S1 window).} ``Which is the tallest building inside the window $\langle$wkt$\rangle$?''
\begin{lstlisting}
CALL spatial.intersects('features', '<wkt>') YIELD node
WITH node WHERE node:Building
  AND node.measured_height IS NOT NULL
RETURN node.id AS id, node.measured_height AS height_m
ORDER BY node.measured_height DESC, node.id ASC LIMIT 1
\end{lstlisting}

\emph{Cross-source (value disagreement).} ``Which buildings have an OSM height that differs from the surveyed height by more than $\langle d \rangle$ meters?''
\begin{lstlisting}
MATCH (b:Building)-[r:ENRICHED_BY {is_primary: true}]
      ->(o:OsmFeature)
WHERE o.osm_height IS NOT NULL
  AND b.measured_height > -999
  AND abs(toFloat(o.osm_height) - b.measured_height) > <d>
RETURN b.id AS id, b.measured_height AS surveyed_m,
       toFloat(o.osm_height) AS osm_m
ORDER BY abs(toFloat(o.osm_height) - b.measured_height)
  DESC, id ASC
\end{lstlisting}

\emph{Provenance (source-restriction trap).} ``Using only authoritative cadastral data, what is the average building height?'' The gold query reads the surveyed property only. A model that also averages the crowd-sourced \texttt{osm\_height} produces a plausibly close but wrong number.
\begin{lstlisting}
MATCH (b:Building)
WHERE b.measured_height > -999
RETURN round(avg(b.measured_height) * 100) / 100.0
       AS avg_height_m
\end{lstlisting}

\emph{Coverage-aware (city-wide count trap).} ``How many buildings in Hamburg have a $\langle$material$\rangle$ roof?'' The gold answer couples the count with its coverage context. A bare count is scored as wrong.
\begin{lstlisting}
MATCH (b:Building)
WITH count(b) AS total,
  count(CASE WHEN b.predictedroofmaterial IS NOT NULL
        THEN 1 END) AS covered,
  count(CASE WHEN b.predictedroofmaterial = '<material>'
        THEN 1 END) AS matching
RETURN matching, covered, total,
  round(1000.0 * covered / total) / 10.0 AS coverage_pct
\end{lstlisting}

\emph{Infeasible (absent attribute).} ``In which year was each building last renovated?'' Gold behavior is an explicit refusal. The guard proves the absence on the target city:
\begin{lstlisting}
MATCH (b:Building)
WHERE b.renovation_year IS NOT NULL
RETURN count(b) AS n  // gate G4: must return n = 0
\end{lstlisting}

\emph{Multilingual (Japanese-key schema bridging, Tokyo).} ``Which buildings belong to the district plan `Ichigaya-Yanagich\=o'?'', also posed in Japanese (\texttt{lang: ja}) as \jp{「地区計画『市谷柳町地区』に属する建物はどれですか。」} The property key itself is Japanese, so the gold query must bridge the schema regardless of the question language:
\begin{lstlisting}[escapeinside={(*@}{@*)}]
MATCH (b:Building)
WHERE b.`(*@\jp{地区計画}@*)` = '(*@\jp{市谷柳町地区}@*)'
RETURN b.id AS id ORDER BY id
\end{lstlisting}

\section{LLM Setup for Text-to-Query}\label{app:full_results_a}

Both query-generation models are evaluated under identical, deliberately \emph{closed-book} conditions. Each question prompt contains only the natural-language question, the project's Cypher and spatial-query rules, and a fixed per-city schema string consisting of the cached APOC-sampled graph schema together with the documented schema-gap patches (\Cref{subsec:experiments_setup}). No live database connection or external tool access is available. The schema string is byte-identical for both models within a given city, ensuring that neither model is exposed to information unavailable to the other. Consequently, a model is expected to refuse requests for attributes that are absent from the provided schema text. Outputs are cached and evaluated offline. Generated Cypher queries are executed against the released graph and compared with gold answers after result-set canonicalization: row multisets are compared with column names removed and floating-point values rounded to $10^{-6}$. As a result, adding an additional column or renaming an existing column alters the row signature and is therefore treated as incorrect rather than silently accepted. 

The model qwen2.5-coder:7b is served locally through Ollama using the \texttt{Q4\_K\_M} quantization, a pinned model tag, greedy decoding (temperature $=0$), the default context window, and a single generation per question. Claude Sonnet~5 is evaluated as a closed-book Claude Code subagent, with the model itself acting as the query generator rather than through the \texttt{ChatAnthropic} API endpoint. No \texttt{ANTHROPIC\_API\_KEY}, live database access, or external tools are available beyond the supplied schema text. The evaluated model snapshot is \texttt{claude-sonnet-5}, tested on 2026-07-19 using the \emph{high} reasoning-effort setting. 

To improve token efficiency, questions within a split are submitted in batches of approximately 25 rather than as individual calls. Each batch prompt explicitly instructs the model to answer every question independently, relying only on the provided schema and rules and without influence from other questions in the batch. This design trades strict per-call isolation for a substantial reduction in the number of model calls. While answering many related questions in a single context may plausibly increase internal consistency relative to fully independent calls, this effect was not measured and no claim is made that the setup is equivalent to one-call-per-question evaluation. 

\section{Representation-Learning Benchmark: Supplementary}\label{app:repL}

\begin{figure*}[t]
  \centering
  \includegraphics[width=\linewidth]{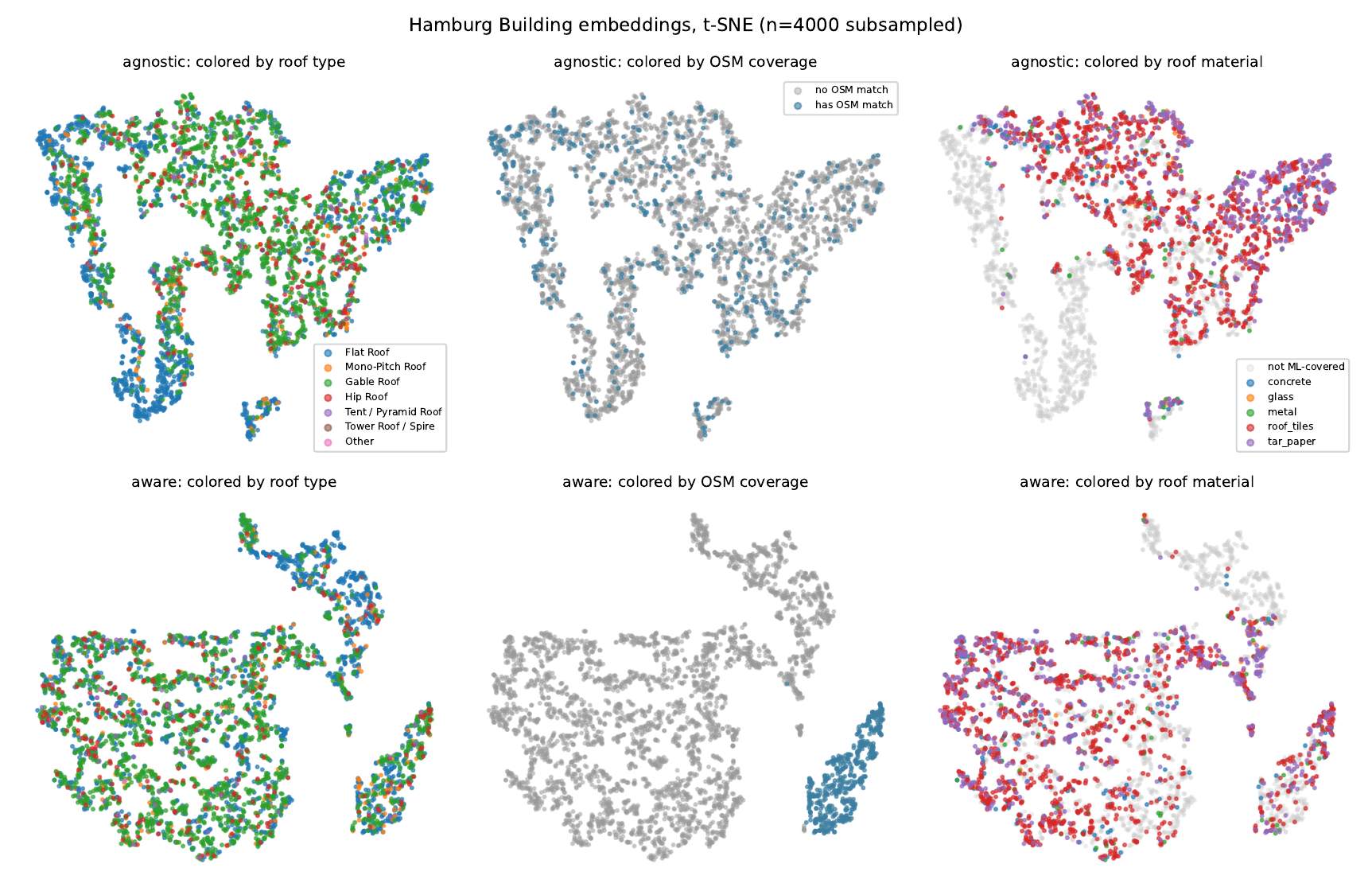}
    \caption{2D t-SNE of Hamburg building embeddings from the T3 matching encoders. Columns show roof type, OSM coverage, and roof-material coverage, where gray marks buildings the ML layer never covered. Compared with the agnostic encoder, the aware encoder more clearly isolates OSM-covered buildings and yields a more structured space.}
  \label{fig:prov-tsne}
\end{figure*}

\Cref{fig:prov-tsne} shows the 2D t-SNE of Hamburg building embeddings from the provenance-agnostic and provenance-aware T3 matching encoders, the qualitative view referenced in \Cref{sec:repL:results}. Table~\ref{tab:repL-crosscity-roof} reports the T2 roof-type cross-city transfer, limited to Hamburg and Helsinki, the only two cities that carry the label.

\begin{table}[ht!]
  \centering
  \caption{T2 roof-type cross-city transfer, leave-one-city-out on the only valid two-city pair (macro-F1 on the merged 3-class label, mean over \num{3} seeds, best per row in bold).}
  \label{tab:repL-crosscity-roof}
  \setlength{\tabcolsep}{4pt}
  \begin{tabular}{@{}lcc@{}}
    \toprule
    Held-out (roof type) & agnostic GNN & aware GNN\\
    \midrule
    Hamburg  & 0.203 & \textbf{0.208}\\
    Helsinki & \textbf{0.336} & \textbf{0.337}\\
    \bottomrule
  \end{tabular}
\end{table}

\end{document}